\documentclass[11pt,fleqn]{article}

\usepackage{a4}
\usepackage{amstext}
\usepackage{amsfonts}
\usepackage{amssymb}
\usepackage{amsmath}
\usepackage{color}
\usepackage{epsfig}
\usepackage{mathtools}
\usepackage{hyperref}

\newcommand{\gtapprox}{\raisebox{-0.5ex}{$\,\stackrel{>}{\scriptstyle\sim}\,$}}

\begin{document}

\begin{center}

{\huge \bf $B\,B$ interactions with static bottom quarks}

{\huge \bf from Lattice QCD}

\vspace{0.5cm}

\textbf{Pedro Bicudo$^a$, Krzysztof Cichy$^{b,c,d}$, Antje Peters$^b$, Marc Wagner$^{b,d}$}

$^a$ Universidade de Lisboa, CFTP, Dep. F\'isica, Instituto Superior T\'ecnico, \\ Av. Rovisco Pais, 1049-001
Lisboa, Portugal\\
$^b$ Goethe-Universit\"at Frankfurt am Main, Institut f\"ur Theoretische Physik,
\\Max-von-Laue-Stra\ss e 1, D-60438 Frankfurt am Main, Germany\\
$^c$ Adam Mickiewicz University, Faculty of Physics, \\Umultowska 85, 61-614 Pozna\'n, Poland\\
$^d$ European Twisted Mass Collaboration

\vspace*{0.3cm}
E-mail: \verb|bicudo@tecnico.ulisboa.pt, kcichy@th.physik-uni-frankfurt.de|, \\
\verb|peters@th.physik-uni-frankfurt.de, mwagner@th.physik-uni-frankfurt.de|

\vspace{0.5cm}

\today

\end{center}

\vspace{0.1cm}

\begin{tabular*}{16cm}{l@{\extracolsep{\fill}}r} \hline \end{tabular*}

\vspace{-0.4cm}
\begin{center} \textbf{Abstract} \end{center}
\vspace{-0.4cm}
The isospin, spin and parity dependent potential of a pair of $B$ mesons is computed using Wilson twisted mass lattice QCD with two flavours of degenerate dynamical quarks.
The $B$ meson is addressed in the static-light approximation, i.e.\ the $b$ quarks are infinitely heavy.
From the results of the $B\,B$ meson-meson potentials, a simple rule can be deduced stating which isospin, spin and parity combinations correspond to attractive and which to repulsive forces.
We provide fits to the ground state potentials in the attractive channels and discuss the potentials in the repulsive and excited channels.
The attractive channels are most important since they can possibly lead to a bound four-quark state, i.e.\ a $\bar{b}\bar{b}ud$ tetraquark.
Using these attractive potentials in the Schr\"odinger equation, we find indication for such a tetraquark state of two static bottom antiquarks and two light $u/d$ quarks with mass extrapolated down to the physical value.

\begin{tabular*}{16cm}{l@{\extracolsep{\fill}}r} \hline \end{tabular*}

\thispagestyle{empty}

\section{Introduction}
The spectrum of hadrons can in principle be computed from Quantum Chromodynamics (QCD). QCD allows for hadrons formed by a quark and an antiquark (i.e.\ mesons -- $q\bar{q}$) and by three quarks (i.e.\ baryons -- $qqq$), as well as for hadrons with more quarks, as long as they form colour singlets.
Prominent examples of the latter are tetraquarks \cite{Jaffe:1976ig,Jaffe:2004ph} and pentaquarks \cite{Lipkin:1987sk,Praszalowicz:2003ik}.
Pentaquarks have recently been observed in the LHCb experiment at the Large Hadron Collider in CERN \cite{Aaij:2015tga}.
However, for tetraquarks, four-quark bound states composed of two quarks and two antiquarks, mostly unconfirmed candidates exist -- e.g.\ light scalar mesons $\sigma$, $\kappa$, $f_0(980)$ and $a_0(980)$, as well as the heavier charm-strange mesons $D_{s0}^\ast$ and $D_{s1}$. 
It is presently not clear whether these states are sufficiently well described by the quark models assuming the standard $q\bar{q}$ structure and it is likely that they are a mixture of this standard quark structure and a tetraquark structure.
Although many of these states, e.g.\ the $Z_c^\pm$, have received considerable experimental attention by the BELLE
collaboration \cite{Liu:2013dau,Chilikin:2014bkk}, the Cleo-C collaboration \cite{Xiao:2013iha}, the
BESIII collaboration
\cite{Ablikim:2013mio,Ablikim:2013emm,Ablikim:2013wzq,Ablikim:2013xfr,Ablikim:2014dxl} and the LHCb
collaboration \cite{Aaij:2014jqa}, it is still necessary to extend these measurements to know more about the decay channels.
Combined with theoretical investigations, it may allow to clarify the status of these candidates in the near future.

The role of lattice QCD in this context is important, since it allows for quantitative predictions directly based on the QCD Lagrangian.
There are several different strategies that can be followed.

First, one can assume the standard quark-antiquark picture and calculate the expected masses of mesons with desired quantum numbers. If the reached precision is conclusive, any deviations of the computed mass from experiment strongly suggest that the underlying $q\bar{q}$ structure is not the correct one.
However, this is at best an indirect vague evidence for a tetraquark structure. 
Moreover, such computations encounter several difficulties, in particular related to decays. 
Furthermore, such investigations are essentially limited to hadrons containing quarks not heavier than the charm quark, since the dynamical treatment of bottom quarks is at present very difficult, i.e.\ it would require very fine lattice spacings.

Second, one can also assume a pure tetraquark structure $qq\bar{q}\bar{q}$, as done e.g.\ in Refs.\ \cite{Prelovsek:2010kg,Alexandrou:2012rm,Mohler:2013rwa,Prelovsek:2014swa,Abdel-Rehim:2014zwa, Berlin:2015faa}.
This is even more challenging, since the number of Wick contractions increases with respect to the $q\bar{q}$ case and it is essential to precisely compute the typically noisy disconnected contributions.
Probably due to these difficulties, Lattice QCD computations with four quarks of finite mass have so far found no evidence for charmed tetraquark bound
states with $cc\bar u \bar d$ \cite{Guerrieri:2014nxa,Ikeda:2013vwa} nor for resonances of the $Z_c$ family
\cite{Prelovsek:2014swa}.

A third possible approach is using the static approximation for all four quarks. Lattice QCD computations show clear evidence for four-body tetraquark potentials 
\cite{Alexandrou:2004ak,Okiharu:2004ve} 
and tetraquark flux tubes 
\cite{Cardoso:2011fq,Cardoso:2012uka}. 
These potentials are then used as input for four-quark models. However, even with a state-of-the-art unitarized four-quark model, the number of tetraquark bound states and resonances produced with the static potentials tend to be too large
\cite{Bicudo:2015bra}. 

The fourth strategy and the one we follow in this work is a compromise between the previously discussed approaches.
We treat the heavy quarks in the static approximation while the light quarks have a finite mass.
This is most appropriate if the heavy quarks are bottom (anti)quarks.
One can then extract via lattice QCD the potentials of two static antiquarks in the presence of two quarks of finite mass.
Here, we extend recent studies of $\bar{b}\bar{b}ud$ tetraquarks \cite{Bicudo:2012qt,Brown:2012tm,Bicudo:2015vta}.
In particular, we consider the light-quark mass dependence of tetraquarks found in Ref. \cite{Bicudo:2015vta} with unphysically heavy $u/d$ quark mass and we find evidence that the binding survives in the limit of the physical pion mass.
Moreover, we discuss in more detail the lattice techniques that we use and we also show complete results for many different $B\,B$ channels, attractive and repulsive and containing ground state as well as excited $B$ mesons.
In the near future, we also plan to perform an extension of our investigations to the related $b \bar{b}$ tetraquarks claimed by the BELLE
Collaboration \cite{Belle:2011aa}. Such tetraquarks are more difficult to study with lattice QCD, since they couple to several decay channels.

\begin{figure}[t!]
\centerline{%
\includegraphics[width=0.85\textwidth]{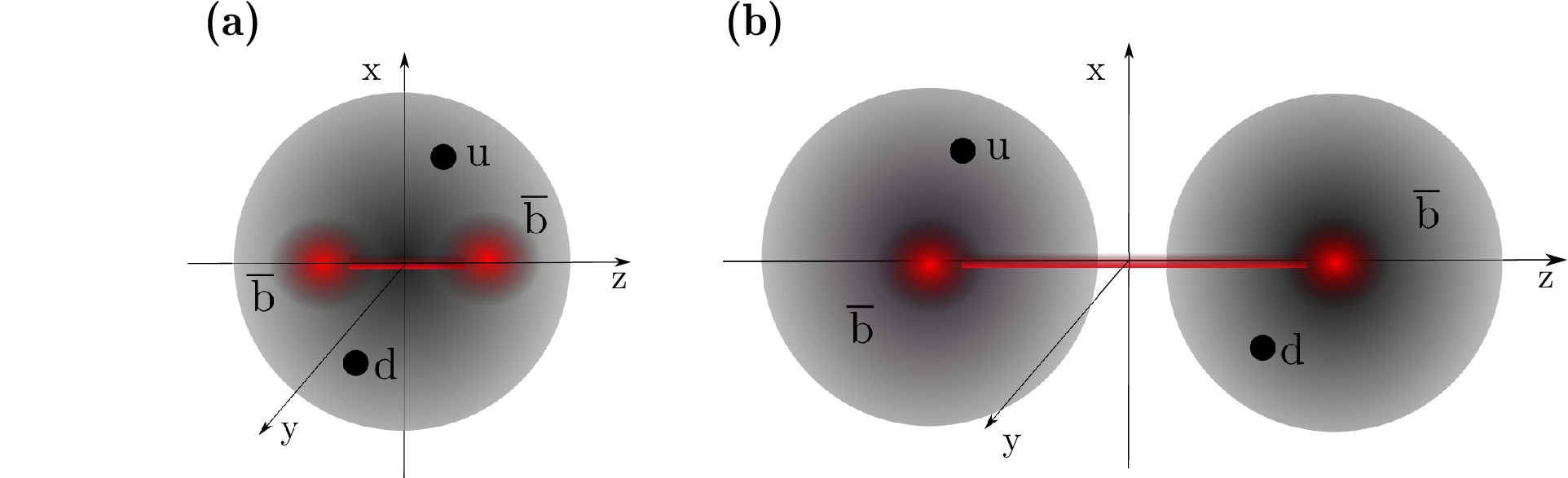}}
\caption{(a) At very short $\bar b \bar b $ separations, the $\bar{b}$ quarks interact with a perturbative one-gluon-exchange Coulomb potential. (b) At large separations the light quarks, for instance $u d$, screen the interaction and the four quarks form two rather weakly interacting $B$ or $B^*$ mesons.
\label{fig:screening}}
\end{figure}

The strategy that we follow to avoid some of the tetraquark technical difficulties follows an idea proposed already in the 1980s 
\cite{Ader:1981db,Ballot:1983iv}. 
We study $\bar{b}\bar{b}ud$ four-quark systems for which it is clear, from the basic principles of QCD, that they form bound states if the antiquarks are heavy enough \cite{Ader:1981db,Ballot:1983iv,Heller:1986bt,Carlson:1987hh,Lipkin:1986dw,Brink:1998as,Gelman:2002wf,Vijande:2003ki,Janc:2004qn,Cohen:2006jg,Vijande:2007ix}. To understand why binding should occur, it is convenient to use the Born-Oppenheimer \cite{Born:1927} perspective, where the wave function of the two heavy antiquarks is determined considering an effective potential describing the light quarks. At very short $\bar b \bar b $ separations, the $\bar{b}$ quarks interact with a perturbative one-gluon-exchange Coulomb potential, while at large separations, the light quarks screen the interaction and the four quarks form two rather weakly interacting $B$ mesons, as illustrated in Fig.\ \ref{fig:screening}. Thus, a screened Coulomb potential is expected. This potential clearly produces a bound state, provided the antiquarks $\bar b \bar b$ are heavy enough.

The calibration problem of quark models is avoided by using $\bar{b} \bar{b}$ potentials obtained from lattice QCD
computations. 
We profit from the fact that the very heavy bottom antiquarks allow for the Born-Oppenheimer approximation \cite{Born:1927}.
From the perspective of the two lighter quarks, the heavy antiquarks $\bar{b} \bar{b}$ can be approximated as static colour charges. 
The static approximation is most appropriate for a comparatively easy extraction of the potential in lattice QCD. 
Then, after the energy of the light quarks $u/d$ is found, it can be used as an effective potential for the heavy antiquarks
$\bar{b} \bar{b}$.
For a detailed introduction to this strategy, see Refs.\ \cite{Bicudo:2012qt,Bicudo:2015vta}. 

Moreover, this approach may turn out to be very interesting for effective models, relying on hadron-hadron potentials. The most important hadron-hadron potential is the $N\,N$ interaction, derived and modelled in great detail by several groups, crucial for our understanding of many aspects of nuclear and astrophysics. A problem with the different $N\,N$ potentials is that they can not be measured directly.
Thus, the lattice QCD potentials between static-light mesons \cite{Richards:1990xf,Mihaly:1996ue,Stewart:1998hk,Michael:1999nq,Cook:2002am,Doi:2006kx,Detmold:2007wk,Bali:2010xa,Wagner:2010ad, Wagner:2011ev}
might be used to test the techniques used by different groups
to derive the $N\,N$ interactions, either from quark degrees of freedom \cite{Ribeiro:1978gx,Myhrer:1987af,Downum:2010qa} 
or from effective degrees of freedom
\cite{Ordonez:1993tn,Kaiser:1997mw,Kaplan:1998tg,Epelbaum:2004fk}.

The paper is organized as follows. In Sec.\ \ref{sec:setup}, we shortly review twisted mass lattice QCD and our lattice setup. Sec.\ \ref{sec:BBcreation} discusses the creation operators used to excite the states that we investigate and the relevant symmetries. Sec.\ \ref{sec:results} presents our numerical results and provides a detailed discussion of our findings. We conclude in Sec.\ \ref{sec:conclusions}.

\section{Twisted mass lattice QCD and lattice setup}
\label{sec:setup}
Twisted mass QCD (tmQCD) is a formulation of QCD with a chirally rotated mass term. It was introduced in
Refs.\ \cite{Frezzotti:2000nk,Frezzotti:2003ni}, where it was shown that this theory is equivalent to
standard QCD, but in its lattice formulation there is a possibility to obtain the so-called
\emph{automatic $\mathcal{O}(a)$ improvement}, which means that the discretization effects are
absent at $\mathcal{O}(a)$, leaving $\mathcal{O}(a^2)$ as the leading cut-off effects. 

In this paper, we work with tmQCD on the lattice (tmLQCD), with $N_f=2$ dynamical quark flavours and the
fermionic action takes
the form \cite{Frezzotti:2000nk,Frezzotti:2003ni}:
\begin{eqnarray}
\label{eq:action}
S_F[\chi,\bar{\chi},U] \ \ = \ a^4 \sum_x \bar{\chi}(x) \Big(D_\mathrm{W} +
i\mu\gamma_5\tau_3\Big) \chi(x),
\end{eqnarray}
where $D_\mathrm{W}$ is the standard Wilson Dirac operator:
\begin{equation}
 D_W = \frac{1}{2} \big( \gamma_{\mu} (\nabla_{\mu} + \nabla^*_{\mu}) - a \nabla^*_{\mu} \nabla_{\mu}
\big),
\end{equation}
with the forward ($\nabla_{\mu}$) and backward ($\nabla^*_{\mu}$) covariant derivatives,
$\chi =
(\chi^{(u)} , \chi^{(d)})$ is the light quark doublet in the so-called \emph{twisted basis}, related
to the physical basis by the twist rotation:
\begin{equation}
\label{eq:twist}
 \psi = e^{i \gamma_5 \tau_3 \omega / 2} \chi
\end{equation}
where $\omega$ is the twist angle. 

The gauge action used in our current work is the tree-level Symanzik improved action
\cite{Weisz:1982zw}:
\begin{equation}
 S_G[U] = \frac{\beta}{3}\sum_x\Big( b_0 \sum_{\mu,\nu=1} \textrm{Re\,Tr} \big( 1 - P^{1\times
1}_{x;\mu,\nu}
\big) 
+ b_1 \sum_{\mu \ne \nu} \textrm{Re\,Tr}\big( 1 - P^{1 \times 2}_{x; \mu, \nu} \big) \Big),
\end{equation}
where $b_1 = -\frac{1}{12}$, $b_0=1-8b_1$, $\beta=6/g_0^2$, $g_0$ is the bare coupling, $P^{1\times
1}$, $P^{1\times 2}$ are the plaquette and the $1\times2$ Wilson loops, respectively. 

To achieve automatic $\mathcal{O}(a)$ improvement, the hopping parameter $\kappa = (8+2 a
m_0)^{-1}$, is tuned to maximal twist by setting it to its critical value, at which the PCAC quark
mass vanishes \cite{Frezzotti:2000nk,Farchioni:2004ma,Farchioni:2004fs,Frezzotti:2005gi,Jansen:2005kk}.
The condition of automatic $\mathcal{O}(a)$ improvement corresponds to tuning $\omega$ in
Eq.\ (\ref{eq:twist}) to $\omega = \pi / 2$.

\begin{table}
\setlength{\tabcolsep}{0.16cm}
\begin{center}
  \caption{\label{tab:setup}Parameters of $N_f=2$ gauge ensembles generated by ETMC
\cite{Boucaud:2007uk,Boucaud:2008xu,Baron:2009wt}. Shown are the inverse bare coupling $\beta$,
lattice size $(L/a)^3\times(T/a)$, bare twisted light sea quark mass in lattice units $a\mu$,
$r_0/a$ \cite{Blossier:2010cr}, lattice spacing $a$ \cite{Baron:2009wt,Jansen:2011vv}, physical
extent of the lattice $L$ in fm and the number of configurations used.}
\begin{tabular}{cccccccc}
    Ensemble & $\beta$ & lattice & $a\mu$ & $r_0/a$ & $a$ [fm] & $L$ [fm] & confs\\
\hline
  B$40.24 $  & 3.90 & $24^3\times48$  & 0.0040 &  5.35(4) & 0.0790(26) &  1.9  & 480 \\
  B$85.24 $  & 3.90 & $24^3\times48$  & 0.0085 &  5.35(4) & 0.0790(26) &  1.9 & 400 \\
  B$150.24 $  & 3.90 & $24^3\times48$  & 0.0150 &  5.35(4) & 0.0790(26) &  1.9 & 260 \\
  \end{tabular}
\end{center}
\end{table}

We use three ensembles of gauge field configurations generated by ETMC, with parameters listed in
Tab.\ \ref{tab:setup}. The lattice spacing is around 0.079 fm and the infinite-volume pion
mass $m_\pi \approx 340$ MeV (B40.24), $m_\pi \approx 480$ MeV (B85.24) and $m_\pi \approx 650$ MeV (B150.24). The
physical spatial extents of the lattices are around 2 fm.
We use a unitary setup, i.e.\ a doublet of degenerate $u/d$ valence quarks described by the action (\ref{eq:action}).
The corresponding $u/d$ quark masses for our ensembles are unphysically heavy, but allow for an extrapolation to the physical light quark masses.

The number of used configurations is 480 for B40.24, 400 for B85.24 and 260 for B150.24, with a
separation of 10 (B40.24) or 20 HMC trajectories (B85.24, B150.24).
On each configuration, $24$ stochastic timeslice propagators are computed ($Z_2 \times
Z_2$ complex noise), $12$ $u$ quark propagators and $12$ $d$ quark propagators, all located on the same
timeslice (this allows to form $\mathcal{O}(12 \times 12)$ samples for $B\,B$ correlators discussed below in detail); the timeslices are cyclically
shifted to further reduce possible autocorrelations.

\section{$B B$ creation operators}
\label{sec:BBcreation}

\subsection{$B$ mesons}
\label{sec:B}
$B$ mesons are mesons containing a $b$ antiquark/quark and a lighter $u$/$d$ quark/antiquark 
(for example, experimentally established mesons included in the PDG review \cite{PDG}: $B^{\pm}$, $B^0$,
$\bar B^0$).
Their masses are around 5.3 GeV and are dominated by the contribution from
the bottom quark/antiquark.
In Lattice QCD, the lighter quarks ($u,d,s,c$) can be treated fully relativistically.
This is, however, not possible when it comes to the bottom quark, whose mass is larger than the typical UV
cut-off on the lattice, i.e.\ the inverse lattice spacing.
One way to treat the $b$ quark is to use Heavy Quark Effective Theory (HQET) \cite{Eichten:1987xu,Eichten:1989zv}.
The leading order of HQET is the static approximation, which means that the bottom quark is treated as
infinitely heavy and hence static, i.e.\ its spatial position is fixed.
This approximation is used in all lattice computations in this paper.

Hence $B$ mesons are approximated as static-light mesons, which are made from a static antiquark
$\bar{Q}$ and either of the light quarks ($u$, $d$)\footnote{The theoretical discussion is very similar in the case of $s$ and $c$ quarks -- see our previous paper \cite{Bicudo:2015vta}.}.
The isospin $I = 1/2$ and $I_z \in \{ -1/2 \, , \, +1/2 \}$.
There are no interactions involving the static quark spin\footnote{However, in nature the static quark spin effects are obviously present in $B$ mesons, since $b$ quarks are not perfectly static. An exploratory study of such effects was presented in Ref.\ \cite{Scheunert:2015pqa} and interesting qualitative results were found.}, hence it is more appropriate to classify static-light mesons by the total angular momentum of their lighter (fully relativistic) degrees of freedom $j$.
The total angular momentum is $j=|l\pm1/2|$, where $l$ is the
orbital angular momentum and $\pm1/2$ is the spin of the light quark. 
We do not consider angular momentum $l\neq0$, therefore $j = 1/2$ and $j_z = \{ -1/2 \,
, \, +1/2 \}$, which is the spin of the $u/d$ quark.
Parity is also a quantum number, $\mathcal{P}\in \{ + \, , \, - \}$.
However, charge conjugation is not a good quantum number, since the two quarks in a meson are
non-identical.

We use yet another common notation to label the states:
\begin{itemize}
\item $S$ denotes the state with $l=0$ ($s$-wave), $j=|j_z|=1/2$, $\mathcal{P}=-$, corresponding to
$B$/$B^\ast$ in Ref.\ \cite{PDG} with $J^\mathcal{P}=0^-$ ($B^\pm$ or $B^0$ mesons) or $J^\mathcal{P}=1^-$ ($B^\ast$); in the static limit,
$B/B^\ast$ mesons are degenerate in mass, since the
parallel or anti-parallel alignment of the spin of the $b$ antiquark and the lighter quark spin does not matter, because
the spin of a static antiquark does not contribute to the energy of the system,
\item $P_-$ denotes\footnote{The notation comes from quark models, in which $P_-$ corresponds to
the orbital angular momentum $l=1$ ($p$-wave). However, in QCD there is (roughly equal) contribution from
both $l=1$ and $l=0$, cf. e.g.\ Ref.\ \cite{Kalinowski:2015bwa}.}
Experimentally, this corresponds to 
$B_0^*$ (with $J^\mathcal{P}=0^+$) or $B_1^*$ ($J^\mathcal{P}=1^+$)
in Refs.\ \cite{Abazov:2007vq,Aaltonen:2007ah} (these are, however, very broad resonances with largely
unknown properties, in particular they are not included in Ref.\ \cite{PDG}).
\item We do not consider angular momentum excitations in this work, but the next states would be $P_+$ ($l=1$
in quark models, $l=1$ or $l=2$ in QCD) with $j=3/2$, $\mathcal{P}=+$, corresponding to
$B_1$
($J^\mathcal{P}=1^+$) or $B_2^*$
($J^\mathcal{P}=2^+$) \cite{Abazov:2007vq,Aaltonen:2007ah} (these resonances are narrow enough and
hence have been much better investigated, in particular $B_1$ and $B_2^*$ are included in
Ref.\ \cite{PDG}), then $D_-$ ($l=2$ in quark models, $l=1$ or $l=2$ in QCD), $j=3/2$, $\mathcal{P}=-$,
etc.
\end{itemize}

The trial states to investigate static-light mesons are $\bar{Q} \Gamma \psi | \Omega \rangle$ for the $\bar{Q}q$ case.
For the ground state ($S$ meson), $\Gamma$ can be chosen as $\gamma_5(1-\gamma_0)/2$, or $\gamma_j(1-\gamma_0)/2$, while for the first excited state ($P_-$ meson), $\Gamma$ can be $(1+\gamma_0)/2$ or $\gamma_j\gamma_5(1+\gamma_0)/2$.
Note that this holds in QCD, while in tmLQCD $S$ and $P_-$ are from the same sector, since parity is not
an exact symmetry (see below).

For a more detailed discussion of static-light mesons, we refer to Refs.\ \cite{Jansen:2008si,Michael:2010aa}.

\subsection{$B B$ systems}
\label{sec:BB}

\subsubsection{Continuum}
Our aim is to to determine the potential of a pair of $B$ mesons as a function of
their spatial separation $r$, taken to be along the $z$ axis.
We will consider static quarks which are antiquarks, i.e.\ the $B$ meson is $\bar{Q}q$. 
Let the positions of the static antiquarks to be $\mathbf{r}_1 = (0,0,-r/2)$ and $\mathbf{r}_2 =
(0,0,+r/2)$, i.e.\ $r=|\mathbf{r}_1-\mathbf{r}_2|$. These coordinates then define the position of each static-light meson.

We now discuss the quantum numbers that characterize the $B\,B$ states. 

\noindent\textbf{Flavour quantum numbers}. The isospin is carried only by the $u$ and $d$ quarks. The $B\,B$ system consisting
of two $\bar{Q}u$ or $\bar{Q}d$ static-light mesons can thus have isospin $I \in \{ 0 \, , \, 1 \}$ and $I_z \in \{ -1 \, ,
\, 0 \, , \,+1 \}$.

\noindent\textbf{Angular momentum}. Rotational symmetry is restricted to rotations around the axis of separation
of static antiquarks, hence the states can be classified by the $z$-component of the total angular
momentum.
Since the spin of the static antiquark plays no dynamical role, it is more appropriate to label states
by $j_z$ of the relativistic $u/d$ quark, i.e.\ $j_z \in \{ -1 \, , \, 0 \, , \, +1 \}$.

\noindent\textbf{Parity}. The states can be labeled by the eigenvalue of the parity operator $\mathcal{P} \in \{ +
\, , \, - \}$.

\noindent\textbf{Reflection across $x$-axis}. For states with $j_z=0$ another symmetry exists -- reflection
around one of the axes perpendicular to the axis of separation, chosen here to be the $x$-axis. We label
the corresponding quantum number by $\mathcal{P}_x$, which can take values $\{ + \, ,\, - \}$.
Note that when using $|j_z|$ instead of $j_z$, $\mathcal{P}_x$ is a quantum number for all states, i.e.\ also for $j_z\neq0$.

Summarizing, $B B$ states can be labeled by five quantum numbers: $(I , I_z , |j_z| , \mathcal{P} ,
\mathcal{P}_x)$.

We now discuss our trial states for $B B$ systems. In general, they take the form:
\begin{eqnarray}
\label{eq:BB}
\mathcal{O}|\Omega\rangle=(\mathcal{C} \Gamma)_{A B} (\mathcal{C}\tilde{\Gamma})_{CD}
\Big(\bar{Q}_C^a(\mathbf{r}_1)
\psi_A^{(f)a}(\mathbf{r}_1)\Big)
\Big(\bar{Q}_D^b(\mathbf{r}_2) \psi_B^{(f')b}(\mathbf{r}_2)\Big) | \Omega \rangle ,
\end{eqnarray}
where $\mathcal{C}$ is the charge conjugation matrix (that can be chosen as $\mathcal{C} = \gamma_0
\gamma_2$), $\Gamma$ and $\tilde{\Gamma}$ are given combinations of $\gamma$ matrices (the possible
choices of $\Gamma$ are listed in Tab.\ \ref{tab:trialstates}, while for $\tilde{\Gamma}$, only
$\tilde{\Gamma}=\{(1-\gamma_0)\gamma_5,\,(1-\gamma_0)\gamma_j\}$, $j=1,2,3$, give non-zero correlators; the obtained
potential does not depend on which $\tilde{\Gamma}$ matrix was chosen), the lower capital
Roman indices are Dirac indices, the upper indices $f,f'$ are flavour indices and $a,b$ are colour
indices.
Note that the coupling of the light degrees of freedom in spinor space via $\Gamma$ determines the quantum numbers
$|j_z|$, $\mathcal{P}$ and $\mathcal{P}_x$.
We consider the following flavour combinations:
\begin{itemize}
 \item $\psi^{(f)} \psi^{(f')} = u d - d u$ with $I=0$,
 \item $\psi^{(f)} \psi^{(f')} = u u$ with $I=1$, $I_z=1$,
 \item $\psi^{(f)} \psi^{(f')} = d d$ with $I=1$, $I_z=-1$,
 \item $\psi^{(f)} \psi^{(f')} = u d + d u$ with $I=1$, $I_z=0$.
\end{itemize}

$B B$ trial states are collected in Tab.\ \ref{tab:trialstates}, together with their quantum numbers.

\begin{table}[t!]
\begin{center}

\begin{tabular}{|c|c||c|c||c|c||}
\hline
\multicolumn{2}{|c||}{\vspace{-0.40cm}} & \multicolumn{2}{c||}{} & \multicolumn{2}{c||}{} \\
\multicolumn{2}{|c||}{} & \multicolumn{2}{c||}{$\psi^{(f)} \psi^{(f')} = u d - d u$} &
\multicolumn{2}{c||}{{$\psi^{(f)}\psi^{(f')}
\in \{ u u \, , \, d d \, , \,u d + d u\}$}} \\
\multicolumn{2}{|c||}{\vspace{-0.40cm}} & \multicolumn{2}{c||}{} & \multicolumn{2}{c||}{} \\
\hline
 & & & & & \vspace{-0.40cm} \\
$\Gamma$ & $|j_z|$ & $\mathcal{P}$, $\mathcal{P}_x$ & result & $\mathcal{P}$, $\mathcal{P}_x$ & result
\\
 & & & & & \vspace{-0.40cm} \\
\hline
 & & & & & \vspace{-0.40cm} \\
$\gamma_5 + \gamma_0\gamma_5$  & $0$ & $-$, $+$ & A, SS & $+$, $+$ & R, SS \\
$1$                          & $0$ & $+$, $-$ & A, SP & $-$, $-$ & R, SP \\
$\gamma_3 \gamma_5$          & $0$ & $+$, $+$ & A, SP & $-$, $+$ & R, SP \\
$\gamma_5 - \gamma_0\gamma_5$  & $0$ & $-$, $+$ & A, PP & $+$, $+$ & R, PP \\
$\gamma_3 + \gamma_0\gamma_3$  & $0$ & $+$, $-$ & R, SS & $-$, $-$ & A, SS \\  
$\gamma_0$                   & $0$ & $-$, $-$ & R, SP & $+$, $-$ & A, SP \\
$\gamma_0 \gamma_3 \gamma_5$ & $0$ & $-$, $+$ & R, SP & $+$, $+$ & A, SP \\
$\gamma_3 - \gamma_0\gamma_3$  & $0$ & $+$, $-$ & R, PP & $-$, $-$ & A, PP \\  
\hline
 & & & & & \vspace{-0.40cm} \\
$\gamma_{1}\gamma_5$          & $1$ & $+$, $-$ & A, SP & $-$, $-$ & R, SP \\
$\gamma_{2}\gamma_5$          & $1$ & $+$, $+$ & A, SP & $-$, $+$ & R, SP \\
$\gamma_2 + \gamma_0\gamma_2$  & $1$ & $+$, $-$ & R, SS & $-$, $-$ & A, SS \\  
$\gamma_1 + \gamma_0\gamma_1$  & $1$ & $+$, $+$ & R, SS & $-$, $+$ & A, SS \\  
$\gamma_0 \gamma_{1} \gamma_5$ & $1$ & $-$, $-$ & R, SP & $+$, $-$ & A, SP \\
$\gamma_0 \gamma_{2} \gamma_5$ & $1$ & $-$, $+$ & R, SP & $+$, $+$ & A, SP \\
$\gamma_2 - \gamma_0\gamma_2$  & $1$ & $+$, $-$ & R, PP & $-$, $-$ & A, PP \\  
$\gamma_1 - \gamma_0\gamma_1$  & $1$ & $+$, $+$ & R, PP & $-$, $+$ & A, PP \\  
\hline
\end{tabular}

\caption{\label{tab:trialstates}Quantum numbers of $B B$ trial states; given are also 
combinations of $\Gamma$ structures that lead to the cancellation of certain states; see also
Tab.\ \ref{tab:mesoncontent}. 
``result'' characterizes the shapes of
the numerically computed $B B$ potentials (A: attractive potential; R: repulsive potential; SS: lower
asymptotic value $2 m(S)$; SP: higher asymptotic value $m(S) + m(P_-)$; PP: highest asymptotic value
$2m(P_-)$).
The states are ordered according to: (1) $|j_z|=0,1$, (2) attractive/repulsive potentials (for the flavour
structure $ud-du$), (3) increasing asymptotic value of the potential, (4) $\mathcal{P}_x=-,+$.}
\end{center}
\end{table}

\subsubsection{Twisted mass lattice QCD}

Working with twisted mass fermions on the lattice, it is convenient to express the trial states in the
twisted basis, related to the physical basis by the chiral rotation (\ref{eq:twist}). The trial states then
read:
\begin{eqnarray}
\label{eq:trialTM}
\mathcal{O}|\Omega\rangle=(\mathcal{C} \Gamma)_{A B} (\mathcal{C}\tilde{\Gamma})_{CD} \Big(\bar{Q}_C^a(\mathbf{r}_1)
\chi_A^{(f_1)a}(\mathbf{r}_1)\Big)
\Big(\bar{Q}_D^b(\mathbf{r}_2) \chi_B^{(f')b}(\mathbf{r}_2)\Big) | \Omega \rangle.
\end{eqnarray}

The lattice formulation of QCD breaks some continuum symmetries that are restored only in the
continuum limit.
Moreover, twisted mass fermions break two additional continuum symmetries (with respect to e.g.\ standard
Wilson fermions) -- parity and isospin.
However, also this breaking is only an $\mathcal{O}(a^2)$ discretization effect (at maximal twist), i.e.
in the continuum these symmetries are restored.
Further below, we discuss the effects of this on the investigated issue, in particular on the labeling of states.

\noindent\textbf{Rotational symmetry}. The space of continuum QCD is symmetric under the rotation group SO(3). On
the lattice, this group is broken to the cubic group H(3), which implies that the symmetry constraints are less
strict and hence mixing within different
representations of the full SO(3) group can occur.
In our case, instead of an infinite number of representations labeled by $j_z = 0, \pm 1, \pm 2, \ldots$,
there are only three different cubic representations, where the continuum representations are mixed, corresponding to $j_z \in \{ 0 , \pm 4, \pm 8 ,
\ldots \}$, to $j_z \in \{ \pm 1 , \pm 3, \pm 5 , \ldots \}$ and to $j_z \in \{ \pm 2 , \pm 6, \pm 10 ,
\ldots \}$.
With our creation operators (\ref{eq:trialTM}), we can study the first two of these cubic
representations. We do not attempt to assign continuum $j_z$ values to the extracted lattice states in a
rigorous way. However, since large angular momentum is usually associated with high energy, it is plausible that the
lowest lying states we investigate correspond to $j_z = 0$ and $|j_z| = 1$, respectively.

\noindent\textbf{Isospin}. 
As we mentioned above, twisted mass lattice QCD breaks isospin at finite lattice spacing.
The most prominent example of this fact is the splitting between the neutral and charged pion masses.
In our investigations, the consequence is that $I$ is not a quantum number, only $I_z$ is conserved. This
leads to a mixing between the continuum sectors $(I=0,\,I_z=0)$ and $(I=1,\,I_z=0)$.\footnote{In principle, also
mixing with $I = 2,3,4,\ldots$ occurs. In practice, however, this is not expected to be problem, as
higher isospin states are related to multi-quark states that have by construction small overlap with our
trial states. Therefore, any mixing with higher isospin states is strongly suppressed.}
As we will mention below, isospin breaking can give some estimate about the size of cut-off effects.

\noindent\textbf{Parity}.
Parity $\mathcal{P}$ is broken by twisted mass fermions.
However, a particular combination of parity and isospin rotation is still a symmetry:
$\mathcal{P}^{(\textrm{tm})}\equiv\mathcal{P}\times[u\leftrightarrow d]$, i.e.\ parity combined with
light flavour exchange.
The properties of trial states under $\mathcal{P}^{(\textrm{tm})}$ depend on the considered flavour
structure:
\begin{itemize}
\item $I_z = 0$ trial states (with light flavour structure $u d \pm d u$) have definite properties under
$\mathcal{P}^{(\textrm{tm})}$ ($u u\leftrightarrow dd$).
\item $I_z = \pm 1$ trial states (light flavour structure $u u$ or $d d$) do not have definite
properties under $\mathcal{P}^{(\textrm{tm})}$.
\item Trial states with light flavour structure $u u \pm d d$ have a definite
$\mathcal{P}^{(\textrm{tm})}$ quantum number, but $I_z$ is not definite.
\end{itemize}
There is no conceptual advantage of using either $uu/dd$ or $uu\pm dd$,
since the spectrum of the two sectors (no matter whether they are split by
$I_z$ or by $\mathcal{P}^{(\textrm{tm})}$) is degenerate. Due to simpler notation, we decide for $uu/dd$.

\noindent\textbf{Reflection around $x$-axis}. As in the continuum, it is important to consider also reflections
around one of the axes perpendicular to the axis of separation. Again, we choose the $x$-axis.
$\mathcal{P}_x^{(\textrm{tm})}$ is defined as $\mathcal{P}_x \times [u\leftrightarrow d]$. The
properties of trial states with different flavour structures are:
\begin{itemize}
\item $I_z = 0$ trial states (light flavour structure $u d \pm d u$) have definite
$\mathcal{P}_x^{(\textrm{tm})}$ properties.
\item For states with $I_z=\pm1$ (light flavour structure $u u$ or $d d$), only
$\mathcal{P}^{(\textrm{tm})} \times \mathcal{P}_x^{(\textrm{tm})}$ is a quantum number.
\end{itemize}

We list all the trial states and quantum numbers for the twisted mass case in two tables: for $I_z = 0$, i.e.\
$\chi^{(1)} \chi^{(2)} = u d \pm d u$, see Tab.\ \ref{tab:ud}, while for $I_z =
\pm1$, i.e.\ $\chi^{(1)} \chi^{(2)} = u u$ or $dd$, see Tab.\ \ref{tab:uu}.

\begin{table}[p!]
\begin{center}
\begin{tabular}{|c|c||c|c|c|c|}
\hline
 & & & & & \vspace{-0.40cm} \\
$\Gamma_X^{(ud \pm du)}$ tb & $\mathcal{P}^{(\textrm{tm})}$, $\mathcal{P}_x^{(\textrm{tm})}$, sec.\ &
$\Gamma_X^{(ud \pm du)}$ pb & $\mathcal{P}$, $\mathcal{P}_x$ & type & mult.\vspace{-0.40cm} \\
 & & & & & \\
\hline
\multicolumn{6}{|c|}{\vspace{-0.40cm}} \\
\hline
\multicolumn{6}{|c|}{\vspace{-0.40cm}} \\
\multicolumn{6}{|c|}{$j_z = 0$, $I = 0$} \\
\multicolumn{6}{|c|}{\vspace{-0.40cm}} \\
\hline
 & & & & & \vspace{-0.40cm} \\
$\gamma_5^{(-)} - i \gamma_0^{(+)}$ & $+$, $-$, $\quad a$ & $(+\gamma_5 + \gamma_0 \gamma_5)^{(-)}$ &
$-$, $+$ & att $S S$ & A \\
$\gamma_0 \gamma_3 \gamma_5^{(-)}$ & $+$, $-$, $\quad a$ & $+\gamma_0 \gamma_3 \gamma_5^{(-)}$ & $-$, $+$
& rep $S P_-$ & A \\
$\gamma_5^{(-)} + i \gamma_0^{(+)}$ & $+$, $-$, $\quad a$ & $(+\gamma_5 - \gamma_0 \gamma_5)^{(-)}$ &
$-$, $+$ & att $P_- P_-$ & A \\
 & & & & & \vspace{-0.40cm} \\
\hline
 & & & & & \vspace{-0.40cm} \\
$\gamma_0 \gamma_3^{(-)} - i \gamma_3 \gamma_5^{(+)}$ & $-$, $+$, $\quad b$ & $(+\gamma_0 \gamma_3 +
\gamma_3)^{(-)}$ & $+$, $-$ & rep $S S$ & B \\
$1^{(-)}$ & $-$, $+$, $\quad b$ & $+1^{(-)}$ & $+$, $-$ & att $S P_-$ & B \\
$\gamma_0 \gamma_3^{(-)} + i \gamma_3 \gamma_5^{(+)}$ & $-$, $+$, $\quad b$ & $(+\gamma_0 \gamma_3 -
\gamma_3)^{(-)}$ & $+$, $-$ & rep $P_- P_-$ & B \\
 & & & & & \vspace{-0.40cm} \\
\hline
 & & & & & \vspace{-0.40cm} \\
$\gamma_3^{(+)}$ & $-$, $-$, $\quad c$ & $+i \gamma_3 \gamma_5^{(-)}$ & $+$, $+$ & att $S P_-$ & C \\
 & & & & & \vspace{-0.40cm} \\
\hline
 & & & & & \vspace{-0.40cm} \\
$\gamma_0 \gamma_5^{(+)}$ & $+$, $+$, $\quad d$ & $+i \gamma_0^{(-)}$ & $-$, $-$ & rep $S P_-$ &
D\vspace{-0.40cm} \\
 & & & & & \\
\hline
\multicolumn{6}{|c|}{\vspace{-0.40cm}} \\
\hline
\multicolumn{6}{|c|}{\vspace{-0.40cm}} \\
\multicolumn{6}{|c|}{$j_z = 0$, $I = 1$, $I_z = 0$} \\
\multicolumn{6}{|c|}{\vspace{-0.40cm}} \\
\hline
 & & & & & \vspace{-0.40cm} \\
$\gamma_0 \gamma_3^{(+)} - i \gamma_3 \gamma_5^{(-)}$ & $-$, $-$, $\quad c$ & $(+\gamma_0 \gamma_3 +
\gamma_3)^{(+)}$ & $-$, $-$ & att $S S$ & E \\
$1^{(+)}$ & $-$, $-$, $\quad c$ & $+1^{(+)}$ & $-$, $-$ & rep $S P_-$ & E \\
$\gamma_0 \gamma_3^{(+)} + i \gamma_3 \gamma_5^{(-)}$ & $-$, $-$, $\quad c$ & $(+\gamma_0 \gamma_3 -
\gamma_3)^{(+)}$ & $-$, $-$ & att $P_- P_-$ & E \\
 & & & & & \vspace{-0.40cm} \\
\hline
 & & & & & \vspace{-0.40cm} \\
$\gamma_5^{(+)} - i \gamma_0^{(-)}$ & $+$, $+$, $\quad d$ & $(+\gamma_5 + \gamma_0 \gamma_5)^{(+)}$ &
$+$, $+$ & rep $S S$ & F \\
$\gamma_0 \gamma_3 \gamma_5^{(+)}$ & $+$, $+$, $\quad d$ & $+\gamma_0 \gamma_3 \gamma_5^{(+)}$ & $+$, $+$
& att $S P_-$ & F \\
$\gamma_5^{(+)} + i \gamma_0^{(-)}$ & $+$, $+$, $\quad d$ & $(+\gamma_5 - \gamma_0 \gamma_5)^{(+)}$ &
$+$, $+$ & rep $P_- P_-$ & F \\
 & & & & & \vspace{-0.40cm} \\
\hline
 & & & & & \vspace{-0.40cm} \\
$\gamma_0 \gamma_5^{(-)}$ & $+$, $-$, $\quad a$ & $+i \gamma_0^{(+)}$ & $+$, $-$ & att $S P_-$ & G \\
 & & & & & \vspace{-0.40cm} \\
\hline
 & & & & & \vspace{-0.40cm} \\
$\gamma_3^{(-)}$ & $-$, $+$, $\quad b$ & $+i \gamma_3 \gamma_5^{(+)}$ & $-$, $+$ & rep $S P_-$ &
H\vspace{-0.40cm} \\
 & & & & & \\
\hline
\multicolumn{6}{|c|}{\vspace{-0.40cm}} \\
\hline
\multicolumn{6}{|c|}{\vspace{-0.40cm}} \\
\multicolumn{6}{|c|}{$j_z = 1$, $I = 0$} \\
\multicolumn{6}{|c|}{\vspace{-0.40cm}} \\
\hline
 & & & & & \vspace{-0.40cm} \\
$\gamma_0 \gamma_{1/2}^{(-)} - i \gamma_{1/2} \gamma_5^{(+)}$ & $-$, $-/+$, $\quad e/f$ & $(+\gamma_0
\gamma_{1/2} + \gamma_{1/2})^{(-)}$ & $+$, $+/-$ & rep $S S$ & I \\
$\gamma_{2/1}^{(+)}$ & $-$, $-/+$, $\quad e/f$ & $+i \gamma_{2/1} \gamma_5^{(-)}$ & $+$, $+/-$ & att $S
P_-$ & I \\
$\gamma_0 \gamma_{1/2}^{(-)} + i \gamma_{1/2} \gamma_5^{(+)}$ & $-$, $-/+$, $\quad e/f$ & $(+\gamma_0
\gamma_{1/2} - \gamma_{1/2})^{(-)}$ & $+$, $+/-$ & rep $P_- P_-$ & I \\
 & & & & & \vspace{-0.40cm} \\
\hline
 & & & & & \vspace{-0.40cm} \\
$\gamma_0 \gamma_{1/2} \gamma_5^{(-)}$ & $+$, $+/-$, $\quad g/h$ & $\gamma_0 \gamma_{1/2} \gamma_5^{(-)}$
& $-$, $-/+$ & rep $S P_-$ & J\vspace{-0.40cm} \\
 & & & & & \\
\hline
\multicolumn{6}{|c|}{\vspace{-0.40cm}} \\
\hline
\multicolumn{6}{|c|}{\vspace{-0.40cm}} \\
\multicolumn{6}{|c|}{$j_z = 1$, $I = 1$, $I_z = 0$} \\
\multicolumn{6}{|c|}{\vspace{-0.40cm}} \\
\hline
 & & & & & \vspace{-0.40cm} \\
$\gamma_0 \gamma_{1/2}^{(+)} - i \gamma_{1/2} \gamma_5^{(-)}$ & $-$, $+/-$, $\quad f/e$ & $(+\gamma_0
\gamma_{1/2} + \gamma_{1/2})^{(+)}$ & $-$, $+/-$ & att $S S$ & K \\
$\gamma_{2/1}^{(-)}$ & $-$, $+/-$, $\quad f/e$ & $+i \gamma_{2/1} \gamma_5^{(+)}$ & $-$, $+/-$ & rep $S
P_-$ & K \\
$\gamma_0 \gamma_{1/2}^{(+)} + i \gamma_{1/2} \gamma_5^{(-)}$ & $-$, $+/-$, $\quad f/e$ & $(+\gamma_0
\gamma_{1/2} - \gamma_{1/2})^{(+)}$ & $-$, $+/-$ & att $P_- P_-$ & K \\
 & & & & & \vspace{-0.40cm} \\
\hline
 & & & & & \vspace{-0.40cm} \\
$\gamma_0 \gamma_{1/2} \gamma_5^{(+)}$ & $+$, $-/+$, $\quad h/g$ & $\gamma_0 \gamma_{1/2} \gamma_5^{(+)}$
& $+$, $-/+$ & att $S P_-$ & L\vspace{-0.40cm} \\
 & & & & & \\
\hline
\end{tabular}
\caption{\label{tab:ud}Twisted basis (tb) and physical basis (pb) quantum numbers for $u d \pm d u$. Different physical basis multiplets are assigned capital letters, while different twisted mass sectors are assigned small letters.}
\end{center}
\end{table}

\begin{table}[t!]
\begin{center}
\begin{tabular}{|c|c||c|c|c|c|}
\hline
 & & & & & \vspace{-0.40cm} \\
$\Gamma_X^{(\frac{uu}{dd})}$ tb & $\mathcal{P}^{(\textrm{tm})} \mathcal{P}_x^{(\textrm{tm})}$, sec.\ &
$\Gamma_X^{(\frac{uu}{dd})}$ pb & $\mathcal{P}$, $\mathcal{P}_x$ & type & mult.\vspace{-0.40cm} \\
 & & & & & \\
\hline
\multicolumn{6}{|c|}{\vspace{-0.40cm}} \\
\hline
\multicolumn{6}{|c|}{\vspace{-0.40cm}} \\
\multicolumn{6}{|c|}{$j_z = 0$, $I = 1$, $I_z = \pm$} \\
\multicolumn{6}{|c|}{\vspace{-0.40cm}} \\
\hline
 & & & & & \vspace{-0.40cm} \\
$\gamma_3 \pm i \gamma_0 \gamma_3 \gamma_5$ & $+$, $\quad i$ & $+\gamma_3 + \gamma_0 \gamma_3$ & $-$, $-$
& att $S S$ & E \\
$\gamma_5$ & $+$, $\quad i$ & $\mp i$ & $-$, $-$ & rep $S P_-$ & E \\
$\gamma_3 \mp i \gamma_0 \gamma_3 \gamma_5$ & $+$, $\quad i$ & $+\gamma_3 - \gamma_0 \gamma_3$ & $-$, $-$
& att $P_- P_-$ & E \\
 & & & & & \vspace{-0.40cm} \\
\hline
 & & & & & \vspace{-0.40cm} \\
$\gamma_0 \gamma_5 \pm i$ & $+$, $\quad i$ & $+\gamma_0 \gamma_5 + \gamma_5$ & $+$, $+$ & rep $S S$ & F
\\
$\gamma_0 \gamma_3$ & $+$, $\quad i$ & $\mp i \gamma_0 \gamma_3 \gamma_5$ & $+$, $+$ & att $S P_-$ & F \\
$\gamma_0 \gamma_5 \mp i$ & $+$, $\quad i$ & $+\gamma_0 \gamma_5 - \gamma_5$ & $+$, $+$ & rep $P_- P_-$ &
F \\
 & & & & & \vspace{-0.40cm} \\
\hline
 & & & & & \vspace{-0.40cm} \\
$\gamma_0$ & $-$, $\quad j$ & $+\gamma_0$ & $+$, $-$ & att $S P_-$ & G \\
 & & & & & \vspace{-0.40cm} \\
\hline
 & & & & & \vspace{-0.40cm} \\
$\gamma_3 \gamma_5$ & $-$, $\quad j$ & $+\gamma_3 \gamma_5$ & $-$, $+$ & rep $S P_-$ & H\vspace{-0.40cm}
\\
 & & & & & \\
\hline
\multicolumn{6}{|c|}{\vspace{-0.40cm}} \\
\hline
\multicolumn{6}{|c|}{\vspace{-0.40cm}} \\
\multicolumn{6}{|c|}{$j_z = 1$, $I = 1$, $I_z = \pm$} \\
\multicolumn{6}{|c|}{\vspace{-0.40cm}} \\
\hline
 & & & & & \vspace{-0.40cm} \\
$\gamma_{1/2} \pm i \gamma_0 \gamma_{1/2} \gamma_5$ & $-/+$, $\quad k/l$ & $+\gamma_{1/2} + \gamma_0
\gamma_{1/2}$ & $-$, $+/-$ & att $S S$ & K \\
$\gamma_{2/1} \gamma_5$ & $-/+$, $\quad k/l$ & $+\gamma_{2/1} \gamma_5$ & $-$, $+/-$ & rep $S P_-$ & K \\
$\gamma_{1/2} \mp i \gamma_0 \gamma_{1/2} \gamma_5$ & $-/+$, $\quad k/l$ & $+\gamma_{1/2} - \gamma_0
\gamma_{1/2}$ & $-$, $+/-$ & att $P_- P_-$ & K \\
 & & & & & \vspace{-0.40cm} \\
\hline
 & & & & & \vspace{-0.40cm} \\
$\gamma_0 \gamma_{1/2}$ & $-/+$, $\quad k/l$ & $\mp i \gamma_0 \gamma_{1/2} \gamma_5$ & $+$, $-/+$ & att
$S P_-$ & L\vspace{-0.40cm} \\
 & & & & & \\
\hline
\end{tabular}
\caption{\label{tab:uu}Twisted basis (tb) and physical basis (pb) quantum numbers for $u u$ and $d d$. Different physical basis multiplets are assigned capital letters, while different twisted mass sectors are assigned small letters.}
\end{center}
\end{table}

\subsubsection{Interpretation of trial states in terms of individual $B$ mesons}
The creation operators in the trial states introduced in Eq.~(\ref{eq:BB}) excite $B\,B$ meson pairs.
The individual $B$ mesons inside these states are, however, not of definite parity and spin.
The $B\,B$ trial states are formed by linear combinations of different $B$ mesons.
To analyze this content, one can introduce parity and spin projectors.
The parity projectors are:
\begin{eqnarray}
P_{\mathcal{P}=\pm} \ \ = \ \ \frac{1 \pm \gamma_0}{2}
\end{eqnarray}
and spin projectors for the non-static quark fields are:
\begin{eqnarray}
P_{j_z = \uparrow,\downarrow} \ \ = \ \ \frac{1 \pm i \gamma_0 \gamma_3 \gamma_5}{2},
\end{eqnarray}
where the plus (minus) sign corresponds to $j_z=\uparrow$ ($j_z=\downarrow$).

We work explicitly in the Dirac representation of the $\gamma$-matrices with the following
conventions:
\begin{eqnarray}
\gamma_0 \ \ = \ \ \left(\begin{array}{cc} 1 & 0 \\ 0 & -1 \end{array}\right) \quad , \quad \gamma_j \
\
= \ \ \left(\begin{array}{cc} 0 & -i \sigma_j \\ +i \sigma_j & 0 \end{array}\right),
\end{eqnarray}
which is the most convenient, since it yields diagonal parity projectors.
The four parity-spin projectors have then the following form:
\begin{eqnarray}
P_{\mathcal{P}=+} P_{j_z = \uparrow} &=& {\rm diag}(1,0,0,0) = (1\,0\,0\,0)\,(1\,0\,0\,0)^T \equiv 
\mathbf{v}_{\mathcal{P}=+ , j_z = \uparrow}^\dagger \mathbf{v}_{\mathcal{P}=+ , j_z = \uparrow},\\
P_{\mathcal{P}=+} P_{j_z = \downarrow} &=& {\rm diag}(0,1,0,0) = (0\,1\,0\,0)\,(0\,1\,0\,0)^T \equiv 
\mathbf{v}_{\mathcal{P}=+ , j_z = \downarrow}^\dagger \mathbf{v}_{\mathcal{P}=+ , j_z = \downarrow},\\
P_{\mathcal{P}=-} P_{j_z = \downarrow} &=& {\rm diag}(0,0,1,0) = (0\,0\,1\,0)\,(0\,0\,1\,0)^T \equiv 
\mathbf{v}_{\mathcal{P}=- , j_z = \downarrow}^\dagger \mathbf{v}_{\mathcal{P}=- , j_z = \downarrow},\\
P_{\mathcal{P}=-} P_{j_z = \uparrow} &=& {\rm diag}(0,0,0,1) = (0\,0\,0\,1)\,(0\,0\,0\,1)^T \equiv 
\mathbf{v}_{\mathcal{P}=- , j_z = \uparrow}^\dagger \mathbf{v}_{\mathcal{P}=- , j_z = \uparrow}.
\end{eqnarray}
The sum of the four above projectors is, of course, the identity operator
\begin{eqnarray}
1 \ \ = \ \ P_{\mathcal{P}=+} P_{j_z = \uparrow} + P_{\mathcal{P}=+} P_{j_z = \downarrow} +
P_{\mathcal{P}=-} P_{j_z = \uparrow} + P_{\mathcal{P}=-} P_{j_z = \downarrow},
\end{eqnarray}
which can be inserted into the light spin coupling of a $B B$ creation operator with a Dirac gamma
structure $\Gamma$:
\begin{eqnarray}
\psi^T \mathcal{C} \Gamma \psi =
\sum_{\substack{\mathcal{P}_1=\pm\\j_1=\uparrow,\downarrow}}
\sum_{\substack{\mathcal{P}_2=\pm\\j_2=\uparrow,\downarrow}}
\psi^T
\mathbf{v}_{\mathcal{P}=\mathcal{P}_1 , j_z = j_1}^\dagger
\underbrace{\mathbf{v}_{\mathcal{P}=\mathcal{P}_1 , j_z = j_1}
 \mathcal{C} \Gamma
 \mathbf{v}_{\mathcal{P}=\mathcal{P}_2 , j_z = j_2}^\dagger}_{=
c_{\mathcal{P}_1 j_1;\mathcal{P}_2 j_2}}
\mathbf{v}_{\mathcal{P}=\mathcal{P}_2 , j_z = j_2} \psi .
\end{eqnarray}
The coefficients $c_{\mathcal{P}_1 j_1;\mathcal{P}_2 j_2}$ represent the static-light meson content,
i.e.\ they can take a value of $\pm1$ or $\pm i$ indicating that a given trial state excites the two $B$
mesons with parity $\mathcal{P}=\mathcal{P}_1$, spin $j_z=j_1$ and parity $\mathcal{P}=\mathcal{P}_2$,
spin $j_z=j_2$ or the value $0$, if a given meson pair is not excited by the considered operator.
We remind that $\mathcal{P}=+$ corresponds to the $P_-$ meson and $\mathcal{P}=-$ to the $S$ meson.
Together with the light cloud angular momentum $j_1$ and $j_2$, there are 16 possibilities for the meson
content related to a given trial state, but only 4 coefficients $c_{\mathcal{P}_1 j_1;\mathcal{P}_2
j_2}$ are always non-zero. The meson contents for all possible $\Gamma$ structures are listed in Tab.\ \ref{tab:mesoncontent}.

\begin{table}[t!]
\begin{center}
\begin{tabular}{|c|c|}
\hline
 & \vspace{-0.40cm} \\
$\Gamma_X$ physical & meson content \\
 & \vspace{-0.40cm} \\
\hline
 & \vspace{-0.40cm} \\
$\gamma_5$ & $- S_\uparrow S_\downarrow + S_\downarrow S_\uparrow - P_\uparrow P_\downarrow +
P_\downarrow P_\uparrow$ \\
$\gamma_0 \gamma_5$ & $- S_\uparrow S_\downarrow + S_\downarrow S_\uparrow + P_\uparrow P_\downarrow -
P_\downarrow P_\uparrow$ \\
$1$ & $- S_\uparrow P_\downarrow + S_\downarrow P_\uparrow - P_\uparrow S_\downarrow + P_\downarrow
S_\uparrow$ \\
$\gamma_0$ & $- S_\uparrow P_\downarrow + S_\downarrow P_\uparrow + P_\uparrow S_\downarrow -
P_\downarrow S_\uparrow$ \\
 & \vspace{-0.40cm} \\
\hline
 & \vspace{-0.40cm} \\
$\gamma_3$ & $-i S_\uparrow S_\downarrow -i S_\downarrow S_\uparrow +i P_\uparrow P_\downarrow +i
P_\downarrow P_\uparrow$ \\
$\gamma_0 \gamma_3$ & $-i S_\uparrow S_\downarrow -i S_\downarrow S_\uparrow -i P_\uparrow P_\downarrow
-i P_\downarrow P_\uparrow$ \\ 
$\gamma_3 \gamma_5$ & $-i S_\uparrow P_\downarrow -i S_\downarrow P_\uparrow +i P_\uparrow S_\downarrow
+i P_\downarrow S_\uparrow$ \\
$\gamma_0 \gamma_3 \gamma_5$ & $-i S_\uparrow P_\downarrow -i S_\downarrow P_\uparrow -i P_\uparrow
S_\downarrow -i P_\downarrow S_\uparrow$ \\
 & \vspace{-0.40cm} \\
\hline
 & \vspace{-0.40cm} \\
$\gamma_1$ & $+i S_\uparrow S_\uparrow -i S_\downarrow S_\downarrow -i P_\uparrow P_\uparrow +i
P_\downarrow P_\downarrow$ \\
$\gamma_0 \gamma_1$ & $+i S_\uparrow S_\uparrow -i S_\downarrow S_\downarrow +i P_\uparrow P_\uparrow -i
P_\downarrow P_\downarrow$ \\ 
$\gamma_1 \gamma_5$ & $+i S_\uparrow P_\uparrow -i S_\downarrow P_\downarrow -i P_\uparrow S_\uparrow +i
P_\downarrow S_\downarrow$ \\
$\gamma_0 \gamma_1 \gamma_5$ & $+i S_\uparrow P_\uparrow -i S_\downarrow P_\downarrow +i P_\uparrow
S_\uparrow -i P_\downarrow S_\downarrow$ \\
 & \vspace{-0.40cm} \\
\hline
 & \vspace{-0.40cm} \\
$\gamma_2$ & $- S_\uparrow S_\uparrow - S_\downarrow S_\downarrow + P_\uparrow P_\uparrow + P_\downarrow
P_\downarrow$ \\
$\gamma_0 \gamma_2$ & $- S_\uparrow S_\uparrow - S_\downarrow S_\downarrow - P_\uparrow P_\uparrow -
P_\downarrow P_\downarrow$ \\ 
$\gamma_2 \gamma_5$ & $- S_\uparrow P_\uparrow - S_\downarrow P_\downarrow + P_\uparrow S_\uparrow +
P_\downarrow S_\downarrow$ \\
$\gamma_0 \gamma_2 \gamma_5$ & $- S_\uparrow P_\uparrow - S_\downarrow P_\downarrow - P_\uparrow
S_\uparrow - P_\downarrow S_\downarrow$\vspace{-0.40cm} \\
 & \\
\hline
\end{tabular}
\caption{\label{tab:mesoncontent}Relation between the physical basis $\gamma$ structure and the
static-light meson content. For brevity, $P_{-;\downarrow/\uparrow}$ is denoted as
$P_{\downarrow/\uparrow}$.}
\end{center}
\end{table}

\section{Numerical results}
\label{sec:results}

\subsection{Lattice techniques}
In this work, we use several techniques to improve the signal quality. 

In particular, to get a better suppression of excited states in correlation functions, we employ operator optimization by APE
smearing \cite{Albanese:1987ds} of spatial links ($N_\textrm{APE} = 30$, $\alpha_\textrm{APE} = 0.5$, cf. Eq.\ (3.4) in Ref.\ \cite{Jansen:2008si}) and
Gaussian smearing \cite{Gusken:1989qx}
($N_\textrm{Gauss} = 50$, $\kappa_\textrm{Gauss} = 0.5$, cf. Eq.\ (3.6) in Ref.\ \cite{Jansen:2008si}).
Moreover, we use stochastic propagators and timeslice
dilution. This part of our setup is very similar to the one used in the study of the static-light meson
spectrum \cite{Jansen:2008si,Michael:2010aa} and baryon spectrum \cite{Wagner:2011fs}.

Similarly to Refs.\ \cite{Jansen:2008si,Michael:2010aa,Wagner:2011fs}, we also use the HYP2 action
for the static quarks (with HYP smearing \cite{Hasenfratz:2001hp} parameters $\alpha_1=\alpha_2=1$,
$\alpha_3=0.5$ \cite{DellaMorte:2005yc}). Below, in Sec.\ \ref{sec:BBfull}, we show a comparison of
results obtained with this action and with the standard Eichen-Hill static action with unsmeared links
representing the world lines of the static quarks.

\subsection{$B\,B$ potentials -- qualitative behaviour}
\label{sec:BBresults}
In this section, we consider the general behaviour of $B\,B$ potentials, in particular we deduce
a rule when a given potential is attractive and when it is repulsive. We use lattice data for our ensemble B40.24 with valence quark mass set to its unitary
value, yielding a pion mass of around 340 MeV. The latter is still unphysically heavy, i.e.\ an extrapolation
to the physical pion mass (see Sec.\ \ref{sec:extrapol}) is needed for solid quantitative statements about e.g.\ the mass differences of the $B$
mesons.

Having a trial state given in Eq.~(\ref{eq:trialTM}), which is of the form $\mathcal{O}|\Omega\rangle$, we can construct the correlation function:
\begin{equation}
\label{eq:corr}
C(t)=\langle\Omega|\mathcal{O}^\dagger(t) \mathcal{O}(0)|\Omega\rangle, 
\end{equation} 
where the argument of $\mathcal{O}$ is Euclidean time.
The effective mass plots for the decay of such
correlation function provide information about the $B\,B$ potential between mesons excited by a given
operator.
For large separations, one expects the saturation of the potentials at a value corresponding to the
sum of masses of the two lightest mesons excited by a given trial state.
When using the elementary $\Gamma$ structures from Tab.\ \ref{tab:mesoncontent}, this can either be two $S$ mesons or one $S$ meson and one
$P_-$ meson, so at large separations the plateau value is $2m(S)$ or $m(S)+m(P_-)$, which differ by
about 400 MeV in the continuum (the value on the lattice can be different due to cut-off effects).
Thus, we also consider linear combinations of $\Gamma$ structures to access states
with asymptotic values of $2m(P_-)$ (cf. Tab.\ \ref{tab:trialstates}) -- an example is $\Gamma=\gamma_5-\gamma_0\gamma_5$, where the
contribution from $SS$ cancels and the one from $P_-P_-$ adds up, such that the ground state
is dominated by the $P_-P_-$ state (otherwise an excited state in both $\Gamma=\gamma_5$ and
$\Gamma=\gamma_0\gamma_5$).

We now consider the number of distinct potentials that can be formed from the considered trial states.
The $B\,B$ system consists of two $B$ mesons and each of such mesons has 2 possibilities for the
quantum numbers $I_z = \pm 1/2$, $j_z = \pm 1/2$, $\mathcal{P} = \pm$, i.e.\ 8 possibilities in total.
This gives $8 \times 8 = 64$ different correlation functions for the $B\,B$ system. Equivalently, one can
consider the spin coupling of the relativistic quarks ($4 \times 4$ matrix $\Gamma$, i.e.\ $16$
possibilities) and their flavour content ($4$ possibilities), which again gives $16\times4=64$ different
correlation functions.

Although some of the potentials have different quantum numbers, they are related in the continuum by
isospin symmetry or by rotational symmetry around the $z$-axis. The former leads to degenerate $I=1$
triplets and the latter to $|j_z|=1$ doublets.
Moreover, there are qualitatively 6 different potentials: attractive potentials with the asymptotic
value of $2m(S)$, $m(S)+m(P_-)$ or $2m(P_-)$ and repulsive potentials with the same asymptotic values.
The number of potentials for each case, together with their degeneracies is:
\begin{center}
\begin{tabular}{llll}
$S S$ potentials,     & attractive: & $1(A) \oplus 3(E) \oplus 6(K)$ & ($10$ states). \\
                      & repulsive:  & $1(B) \oplus 3(F) \oplus 2(I)$ & ($\phantom{0}6$ states).
\vspace{0.1cm} \\
$S P_-$ potentials,   & attractive: & $1(B) \oplus 1(C) \oplus 3(E) \oplus 3(G) \oplus 2(I) \oplus 6(L)$
& ($16$ states). \\
                      & repulsive:  & $1(A) \oplus 1(D) \oplus 3(F) \oplus 3(H) \oplus 2(J) \oplus 6(K)$
& ($16$ states). \vspace{0.1cm} \\
$P_- P_-$ potentials, & attractive: & $1(A) \oplus 3(E) \oplus 6(K)$ & ($10$ states). \\
                      & repulsive:  & $1(B) \oplus 3(F) \oplus 2(I)$ & ($\phantom{0}6$ states),
\vspace{0.2cm}
\end{tabular} \\
\end{center}
where we have used multiplet labelling by $A,B,C,\ldots$ from Tabs.~\ref{tab:ud} and \ref{tab:uu}.
Thus, the $64$ trial states (\ref{eq:BB}) lead to $24$ different potentials in the continuum.
However, due to isospin symmetry breaking in twisted mass LQCD, the $I=1$ potentials with $I_z=0$ and $I_z=\pm1$ and otherwise identical quantum numbers are not exactly degenerate. Since isospin breaking is a discretization effect, these states
become degenerate in the continuum limit. In this way, the difference between them can be considered as
a crude estimate of the magnitude of cut-off effects.

\begin{figure}[t!]
\begin{center}
\includegraphics[width=0.345\textwidth,angle=270]{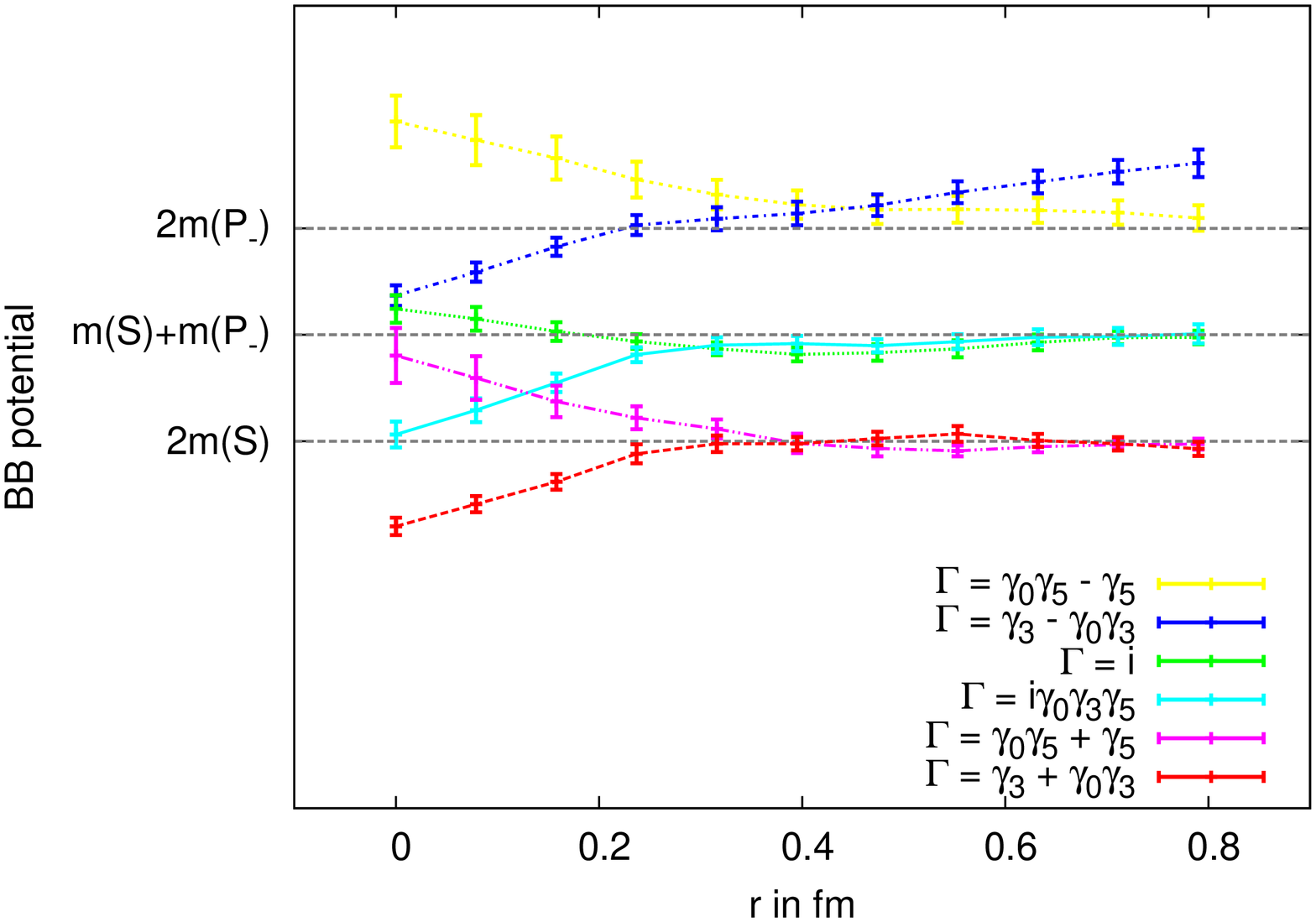}
\includegraphics[width=0.345\textwidth,angle=270]{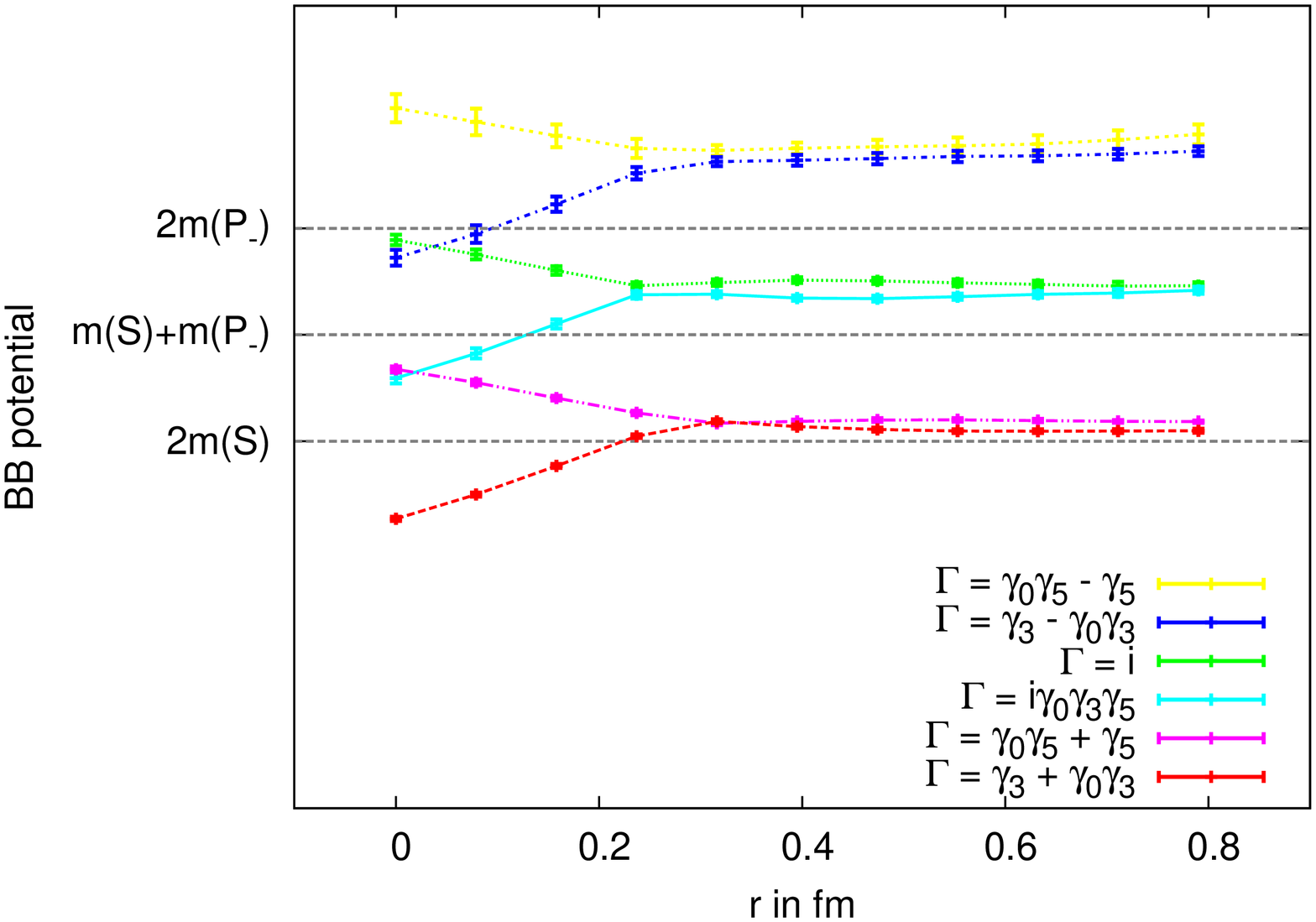}
\caption{\label{fig:examples}Examples of 6 types of $B B$ potentials as functions of the separation
$r$: attractive and repulsive, with 3 different asymptotic values: $2m(S)$, $m(S)+m(P_-)$ or $2m(P_-)$, (left) from 6 individual correlation functions, (right) from a $6\times6$ correlation matrix.
The flavour structure is $uu$.}
\end{center}
\end{figure}

As we already mentioned, there are qualitatively 6 different cases, attractive/repulsive potentials with
3 different asymptotic values, depending on the lowest state excited by the given creation
operator.
Examples for all these cases are shown in Fig.\ \ref{fig:examples}.
The correlation functions entering this plot are members of the same twisted mass sector (called ``$i$'' in Tab.\ \ref{tab:uu}).
In the left plot, we show the potentials extracted from single correlators, while in the right plot, the potentials are extracted from a $6\times6$ correlation matrix using the generalized eigenvalue problem \cite{Blossier:2009kd}.
The horizontal lines correspond to the 3 different asymptotic values, as extracted from a $2\times2$ correlation matrix of 2 static-light meson correlation functions.
In both plots, we obtain qualitative agreement with the expectations from Tab.\ \ref{tab:uu} about the behaviour of potentials.
Differences within errors are observed only for the cases of $SP_-$ and $P_-P_-$ potentials and are probably due to contamination by excited states (short plateaus at small $t$ because of limited signal quality).
However, for $SS$ potentials, there are no such differences and hence there is no problem for our bound states analysis.

It is important to emphasize that even if the quantum numbers of considered potentials are the same, the
resulting types of potential can be different.
In the case of Fig.\ \ref{fig:examples}, all quantum numbers are the same in the twisted basis, but we observe both attractive and repulsive potentials and all three possible asymptotic values.
This results from the fact that, as Tab.\ \ref{tab:mesoncontent} indicates, different cases have very different meson contents.
Thus, the states that are excited for different correlators with the same quantum numbers can be very different -- in some cases it can mainly be the ground state and in some cases mainly some excited state.

The complete analysis of 36 independent potentials in the twisted mass case leads to the observations about whether a potential is attractive or repulsive, reported in Tabs.\ \ref{tab:ud} and \ref{tab:uu}.
They can be summarized in terms of a simple rule that involves the behaviour of the trial state corresponding to the potential under meson exchange,
i.e.\ under combined exchange of flavour, spin and parity.
\emph{If the trial state is symmetric/antisymmetric under such meson exchange, the resulting potential is then
attractive/repulsive.}
Consideration of the individual meson content for the trial states of potentials included in
Fig.\ \ref{fig:examples} (and Figs.\ \ref{FIG003_j0_I0} and \ref{FIG003_j0_I1} below) allows to easily verify this rule.
The deduced rule is in agreement with a less general rule obtained in earlier quenched computations of $B\,B$ potentials
\cite{Michael:1999nq,Detmold:2007wk}, which, however, only considered potentials between ground state $S$ mesons.
It can also be justified in the following way.
Consider the one-gluon exchange between two static antiquarks $\bar{Q}\bar{Q}$. The resulting potential, valid for small $r$,
is of Coulomb type $V(r)\propto g^2/r$, where $g$ is the coupling and $r$ is the separation between
antiquarks. The $\bar{Q}\bar{Q}$ pair can be either in the colour $\bf{3}$ or $\bar{\bf{6}}$
representation. For the former case, the proportionality constant is negative (attractive potential) and
for the latter positive (repulsive potential) due to the Casimir scaling.
Let us now consider the attractive case. 
The light quark wave functions have to be antisymmetric (Pauli principle).
If the static antiquarks are in the $\bf{3}$ representation, the light quarks are in $\bar{\bf{3}}$,
which is antisymmetric. This implies that the symmetry with respect to combined exchange of flavour, spin
and parity of the light quarks can not alter the total antisymmetry of the wave function, i.e.\ the trial
state has to be symmetric under such combined exchange.
Analogously, the repulsive symmetric $\bar{\bf{6}}$ representation together with the antisymmetry
of the fermionic wave function implies that the repulsive potential is obtained when the trial
state is antisymmetric with respect to the above mentioned combined exchange.

\subsection{$B\,B$ potentials -- results for all channels}
\label{sec:BBfull}
\begin{figure}[t!]
\begin{center}
\includegraphics[width=0.345\textwidth,angle=270]{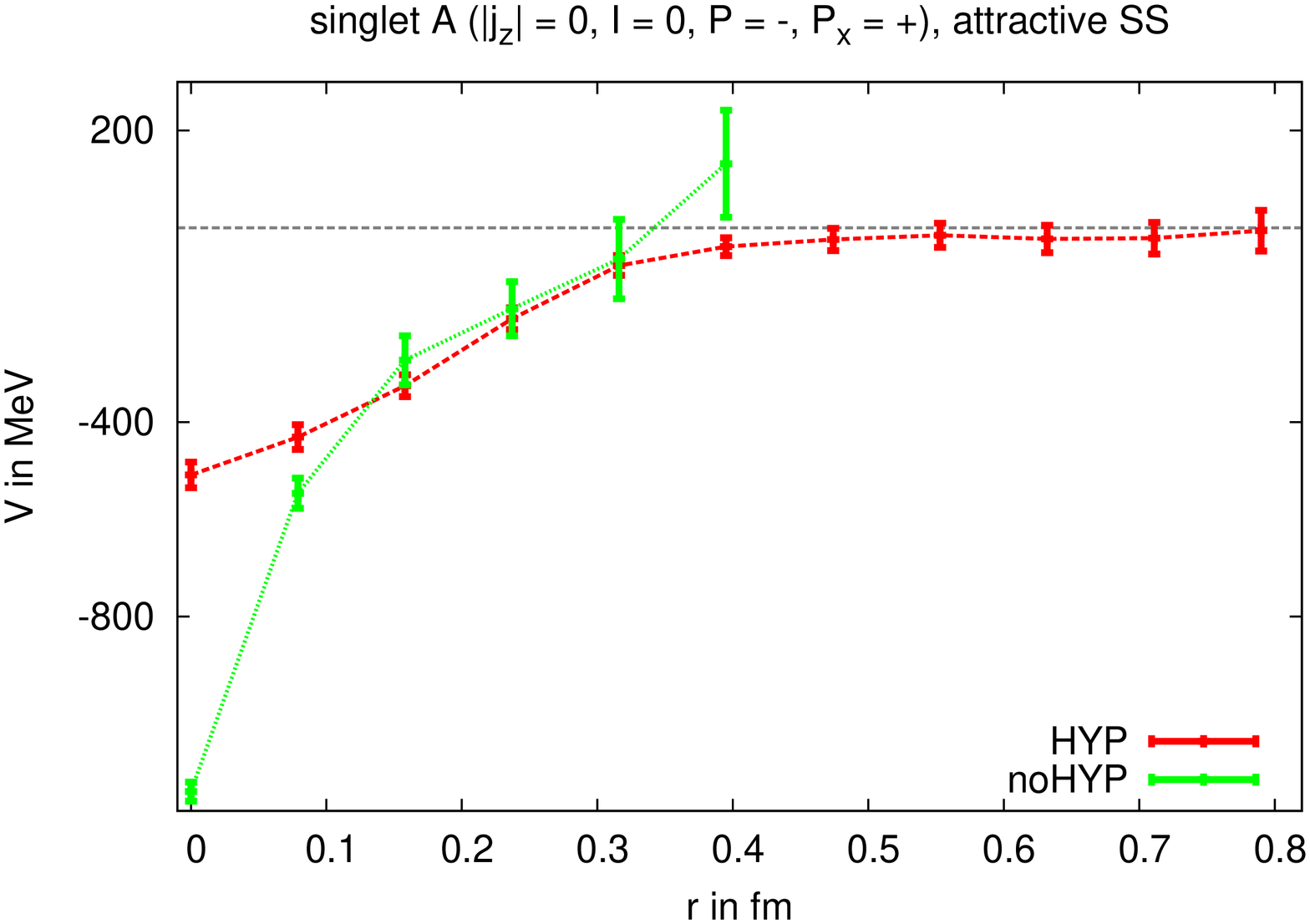}
\includegraphics[width=0.345\textwidth,angle=270]{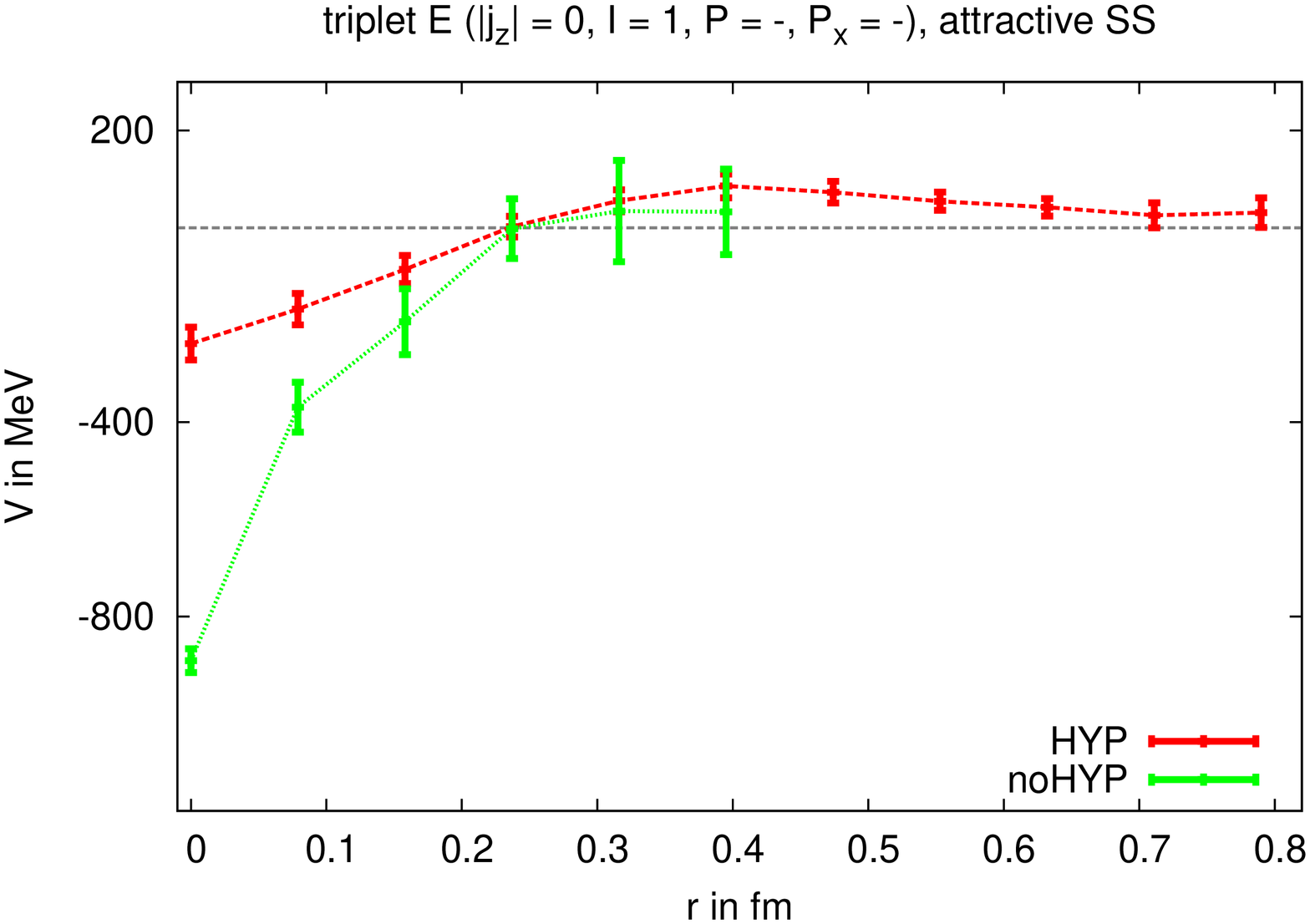}
\includegraphics[width=0.345\textwidth,angle=270]{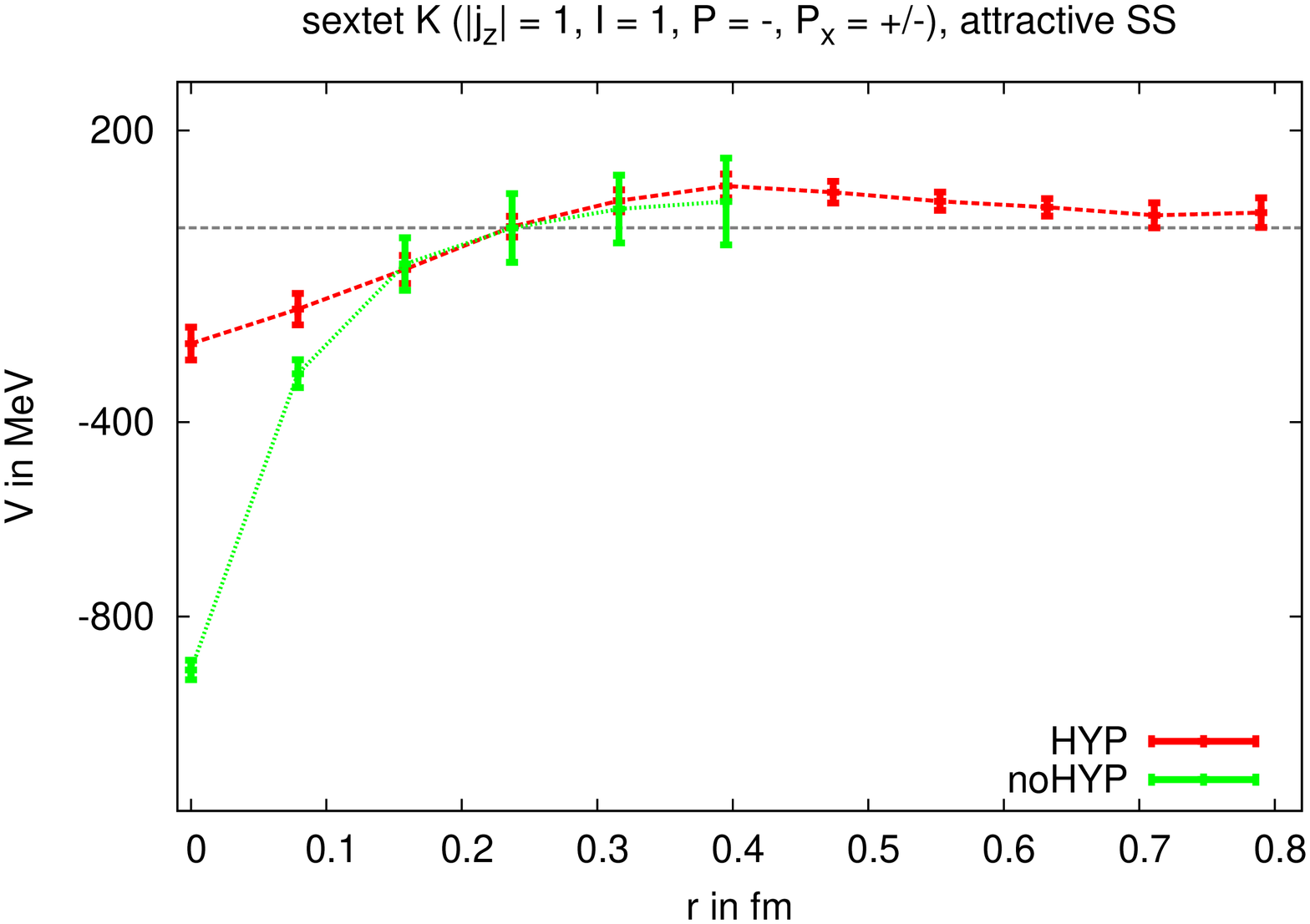}
\caption{\label{fig:HYP}Comparison of attractive $SS$ potentials obtained with HYP2 smearing (red solid lines) and without HYP2
smearing (green dashed lines). Shown are only attractive potentials with $2m(S)$ as the asymptotic
value (subtracted from the potential), from multiplets A, E and K of Tabs.~\ref{tab:ud},~\ref{tab:uu}.}
\end{center}
\end{figure}

In this subsection, we again use lattice data
from the B40.24 ensemble.
We start by comparing results with and without HYP smearing of world lines of static quarks -- see Fig.\ \ref{fig:HYP}.
We show only attractive potentials with $2m(S)$ (subtracted from the potentials) as the asymptotic value, coming from multiplets A, E and
K of Tabs.\ \ref{tab:ud} and \ref{tab:uu}.
In general, for separations $r\gtapprox2a$, the results from HYP and non-HYP cases are compatible, but
with visibly reduced statistical errors for the former. In the small-separation region, however, HYP
smearing significantly distorts the potential by filtering out UV fluctuations.
We, therefore, restrict the remaining discussion to separations $r\geq2a$ and use the HYP results, which exhibit significantly smaller errors.

\begin{figure}[t!]
\begin{center}
\input{FIG003_j0_I0.pstex_t}
\caption{\label{FIG003_j0_I0} All 12 extracted potentials in the $I=0$ channel. From all potentials the asymptotic value of $2m(S)$ is subtracted. The three horizontal lines correspond to the asymptotic values of $2m(S)$, $m(S)+m(P_-)$ and $2m(P_-)$.}
\end{center}
\end{figure}

\begin{figure}[t!]
\begin{center}
\input{FIG003_j0_I1.pstex_t}
\caption{\label{FIG003_j0_I1}  All 24 extracted potentials in the $I=1$ channel. From all potentials the asymptotic value of $2m(S)$ is subtracted. The three horizontal lines correspond to the asymptotic values of $2m(S)$, $m(S)+m(P_-)$ and $2m(P_-)$.}
\end{center}
\end{figure}

We present our results for all ($I$, $I_z$, $|j_z|, \mathcal{P}, \mathcal{P}_x$) channels in Figs.\ \ref{FIG003_j0_I0} and \ref{FIG003_j0_I1}, including for the first time also potentials between excited mesons. 
The potentials were shifted vertically, such that the zero value corresponds to the
asymptotic value of $2m(S)$.
The horizontal lines correspond to $2 m(S)$, $m(S) + m(P_-)$ and $2 m(P_-)$.

Fig.\ \ref{FIG003_j0_I0} shows 12 potentials corresponding to $I=0$, with flavour structure $ud\pm du$,
quantum numbers contained in the label of each plot and $\Gamma$-structures given in Tab.\ \ref{tab:ud}.
There are four singlet channels ($I=0$, $j_z=0$) with different combinations of
$\mathcal{P}/\mathcal{P}_x$ and two spin doublet channels ($I=0$, $j_z=\pm1$) with either
$\mathcal{P}=+$ or $\mathcal{P}=-$.
One of the main physical motivations of this paper is to find cases where it is likely that two $B$
mesons can form a bound tetraquark state.
From this point of view, attractive potentials with an asymptotic value of $2m(S)$ are the
most interesting -- for bound state formation an attractive potential is mandatory, while restriction to
the ground state allows to obtain the best quality of data.
Moreover, excited states would require a very strongly bound state, otherwise the decay to a ground state $SS$ pair is possible and a statement about binding is not very strong. 
Hence, the spin/isospin singlet A (scalar isosinglet) with $\Gamma=\gamma_5+\gamma_0\gamma_5$ and flavour
structure $ud-du$
is the natural candidate for further phenomenological investigations,
reported already in our previous publication \cite{Bicudo:2015vta} and extended below.
In Sec.\ \ref{sec:extrapol}, we check whether the binding previously observed at an unphysically heavy pion mass survives or is even enhanced by the physical pion mass limit.

In the plots for $I = 1$ (Fig.\ \ref{FIG003_j0_I1}, 24 potentials, flavour structure $uu$, $dd$ or
$ud+du$), solid lines correspond to $I_z = 0$, while the dashed ones to $I_z = \pm 1$.
There are four isospin triplet channels ($I=1$, $j_z=0$) with different combinations of
$\mathcal{P}/\mathcal{P}_x$ and two spin/isospin sextet channels ($I=1$, $j_z=\pm1$) with either
$\mathcal{P}=+$ or $\mathcal{P}=-$.
The difference between $I_z=0$ and $I_z=\pm1$ is due to the breaking of isospin symmetry by twisted mass
fermions. As such, it is only a discretization effect and actually the differences in the results for
these two $I_z$ channels can be used to estimate the size of cut-off effects in our computations.
This is particularly important, since we have only used one lattice spacing at the moment.
As can be seen in Fig.\ \ref{FIG003_j0_I1}, the above mentioned differences between $I_z$ channels give
consistent results in most of the cases (i.e.\ within statistical error of each other). Only in few
cases the differences exceed the $1\sigma$-level, but they are never larger than approximately $2\sigma$.
This allows us to conclude that the discretization effects are rather small in our setup and justifies the use of these results to conclude about the behaviour in the continuum.
As mentioned above, we are especially interested in attractive potentials with the asymptotic value of
$2m(S)$.

\subsection{Using $B\,B$ potentials in the Schr\"odinger equation}
The aforementioned scalar isosinglet A can be treated in a relatively simple way. It has light spin $j=0$ and hence the corresponding potential is spin-independent. Therefore, one can use the lattice potential $V(r)$ in the radial Schr\"odinger equation:
\begin{equation}
\label{eq:schr}
 \left(-\frac{\hbar^2}{2m_H} + V(r) \right) S(r) = ES(r),
\end{equation}
where $m_H$ is set to either the bottom quark mass or the $B$ meson mass, $S(r)=R(r)/r$ and $R(r)$ is the radial part of the full wave function, $\psi(r,\theta,\phi)=R(r)Y_{lm}(\theta,\phi)$ ($Y_{lm}$ denote spherical harmonics).

The $I=1$ channels provide further examples of attractive potentials with the asymptotic value of $2m(S)$ -- in the triplet E (with $I_z=0$ and
$|I_z|=1$) and in the sextet K (also for both $I_z=0$ and $|I_z|=1$), with
$\Gamma=\gamma_j+\gamma_0\gamma_j$ ($j=1,2,3$).
These multiplets have light spin $j=1$, E corresponds to $j_z=0$, while K to $j_z=\pm1$, and consequently the potential which has to be used in the Schr\"odinger equation is spin-dependent.
The Schr\"odinger equation then reads:
\begin{equation}
\left(-\frac{\hbar^2}{2m_H}1_{\rm spin}\Delta + \left(\mathbf{e}_r\otimes\mathbf{e}_r V_E(r)+\Big(\mathbf{e}_\theta\otimes\mathbf{e}_\theta+\mathbf{e}_\phi\otimes\mathbf{e}_\phi\right)V_K(r) \Big)\right)\vec\psi(\mathbf{r}) = E\vec\psi(\mathbf{r}),
\end{equation}
where the $z$-component of $\vec\psi$ is the wave function of the $j_z=0$ state, while the $x$- and $y$-components correspond to $j_x=0$ and $j_y=0$ or (taking suitable linear combination)  $j_z=\pm1$.
Clearly, this is a much more complicated problem than for the multiplet A, because here we have 3 coupled differential equations.
In practice, however, the potentials $V_E$ and $V_K$ are identical within our numerical precision, i.e.\ $V_E=V_K$, and using this equality in the above equation reduces it to a form analogous to Eq.~(\ref{eq:schr}).
Still, the potentials obtained for the E and K case are  much less
deep and broad (i.e.\ less attractive) than the one in the isospin singlet A and hence do not lead to the formation of a bound state.
As such, we do not discuss them further.

The approach of plugging the lattice extracted potentials into the Schr\"odinger equation has been extensively used in our previous publications \cite{Bicudo:2012qt,Bicudo:2015vta}. Here, we summarize its main outcomes, obtained using two ensembles of gauge field configurations -- B40.24 and an ensemble with roughly twice smaller lattice spacing (called E17.32). Both ensembles correspond to a single non-physical pion mass of around 340 MeV.

On the B40.24 configurations, we investigated a unitary setup of static-light mesons built of two static antiquarks and two light, but unphysically heavy, $u$/$d$ quarks. We analyzed in detail the binding in two isospin channels, yielding attractive potentials with $2m(S)$ as the asymptotic value -- scalar isosinglet and vector isotriplet. Using the Schr\"odinger equation, we found clear indication for binding in the scalar isosinglet case. For the vector isosinglet, we found no binding.
It is worth to emphasize that these results correspond to a non-physical value of the pion mass. The theoretical expectation is that binding increases towards the physical pion mass. Therefore, it might be the case that the vector isotriplet case can also be binding at the physical values of the $u/d$ quark masses.

On the E17.32 ensemble, we considered a partially quenched setup of static-strange and static-charm mesons -- the valence quark masses were set to either the physical strange or the physical charm quark mass. For this case, the lattice spacing of B40.24 was not fine enough, mostly since heavier non-static quarks lead to narrower potentials, and hence a twice smaller lattice spacing was used. Analyzing the potentials, we also found no evidence for binding and we do not expect that a lower sea quark mass will change this result.

In our present paper, we now have the results for three different pion masses for the unitary setup. We can therefore address the qualitative question whether $\bar{b}\bar{b}ud$ can form a bound state at physically light $u/d$ quark masses and also quantitatively determine the corresponding binding energy.

\subsection{$B\,B$ potentials at the physical pion mass}
\label{sec:extrapol}
All the results that we have shown so far correspond to only a single ensemble of gauge field configurations -- B40.24, at a lattice spacing $a\approx0.079$ fm and at a pion mass of around 340 MeV.
Finally, we are of course interested in the continuum results at the physical pion mass.
We have argued above that cut-off effects are comparatively small in our setup by comparing lattice results corresponding to the same continuum channel, but affected by different discretization effects -- we have found that the results from these different lattice channels are compatible within statistical uncertainties.

However, the question remains to what extent the unphysically heavy $u/d$ quark mass affects our conclusion -- in particular how much stronger the binding is at physically light $u/d$ quark masses for the scalar isosinglet and whether binding occurs for the vector isotriplet case.
To address this question, we performed computations for two additional ensembles corresponding to the same lattice spacing -- B85.24 and B150.24 with pion masses of around 480 MeV and 650 MeV, respectively.
We use the same strategy as in Ref.\ \cite{Bicudo:2015vta}, i.e.\ to quantify systematic errors we perform several fits of the potentials in different ranges $t_{\rm min}\leq t \leq t_{\rm max}$ of temporal separations of the correlation function $C(t,r)$ at which the potential $V(r)$ is read off and different ranges of the static quark separation $r_{\rm min}\leq r \leq r_{\rm max}$ of the potential $V(r)$.
The fitting ansatz is:
\begin{equation}
V(r)=-\frac{\alpha}{r}\exp\left(-\left(\frac{r}{d}\right)^p\right)+V_0, 
\label{eq:ansatz}
\end{equation}
where the fitting parameters are $\alpha$, $d$ and $V_0$, while $p$ is fixed to 2 from phenomenological considerations.

Notice the fit of the $B \, B$ potential in Eq.\ (\ref{eq:ansatz}) is also interesting for effective hadronic models, which assume hadron-hadron potentials, and welcome lattice QCD data to compare with. In particular, in the literature $D \, D$ and $B \, B$ interactions are computed from meson exchange diagrams
\cite{Xiao:2013yca,Ozpineci:2013qza,Torres:2013lka,Aceti:2014kja,Aceti:2014uea,Dias:2014pva}, and are applied to the study of the $Z_c^\pm$ and $Z_b^\pm$ tetraquarks. Moreover, there are also computations  
of the $B \, B$ interactions in the open bottom channels \cite{Barnes:1999hs} similar to ours, including the prediction of binding into a tetraquark.\footnote{ In the closed bottom $B \, B$ channels, the effective hadron models produce $B \, B$ interactions with terms at different ranges, for instance the one-pion exchange potential is a long range potential, while the one-$\rho$ exchange potential is a short range potential. When the long range and short range potentials have different signs, the overall potential may have a dual behaviour,  for instance repulsive at short separations and attractive at long separations, as it occurs in the well known $ N \, N$ interaction. However, in our open bottom case, we have not found any clear evidence of such a dual behaviour. Our $B\,B$ potentials are either attractive or repulsive, consistent with the colour screening model. 
}

The expressions for the potentials obtained from the fits are then plugged into the Schr\"odinger equation and for each fit the binding energy $E_B$ is computed. If $E_B<0$, we conclude that the given potential leads to binding of the interacting $B$ mesons.
Having several fitting ranges allows us to reliably investigate the systematic uncertainties related to the choice of the $t$- and $r$-intervals.
To this aim, we build distributions of the fitting parameters $\alpha$, $d$ and of the binding energy $E_B$.

\begin{figure}[t!]
\begin{center}
\includegraphics[width=0.7\columnwidth,angle=270]{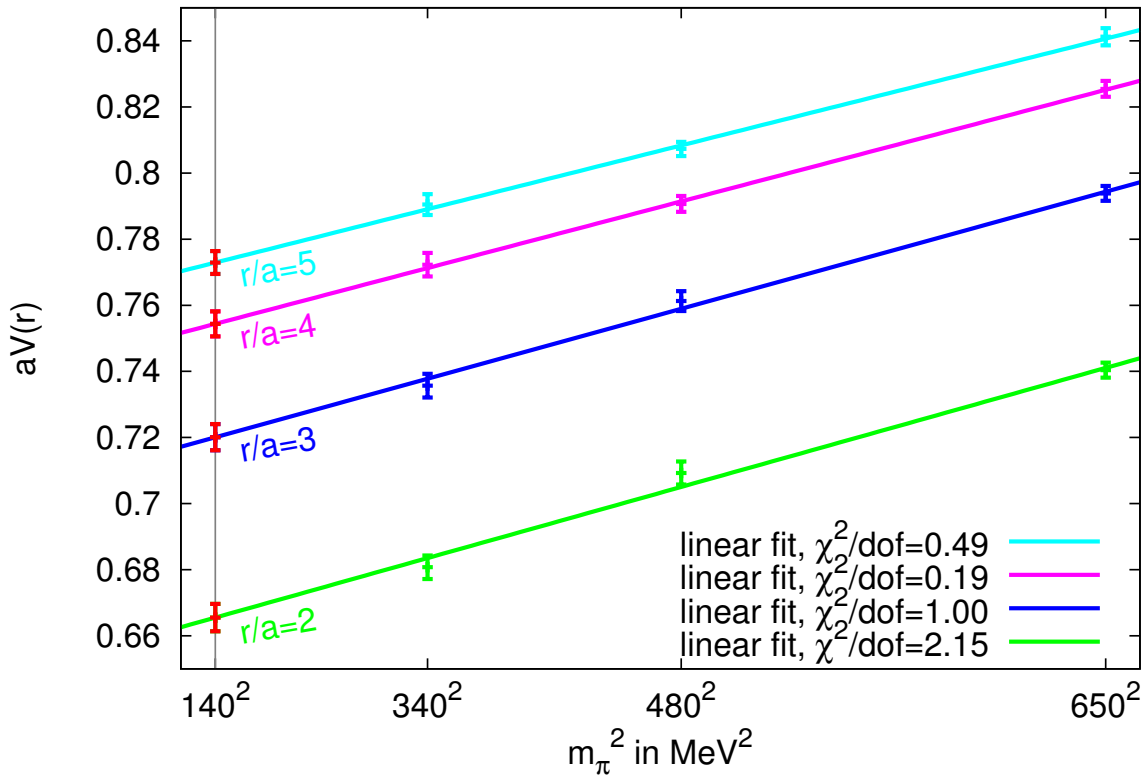}
\caption{\label{fig:Vextrapol} Examples of fits of Eq.~(\ref{eq:Vextrapol}) for the $t$-interval $t/a\in[4,9]$ and for $r/a=2,3,4,5$ in the scalar isosinglet case. Three pion masses are used to extrapolate to the physical pion mass.}
\end{center}
\end{figure}

Moreover, for each $t$- and $r$-range, we also extrapolate the potentials to the physical pion mass, using a linear ansatz in $m_\pi^2$, i.e.\ 
\begin{equation}
\label{eq:Vextrapol}
V(r,m_\pi)=V(r,m_\pi^{\rm phys})+c(m_\pi^2-(m_\pi^{\rm phys})^2), 
\end{equation}
where $V(r,m_\pi^{\rm phys})$ and $c$ are fitting parameters and $m_\pi^{\rm phys}$ is the physical pion mass.
Examples of such fits are shown in Fig.\ \ref{fig:Vextrapol} for the $t$-interval $t/a\in[4,9]$ and for $r/a=2,3,4,5$.
In all cases, the linear fitting ansatz gives good description of lattice data.
The extrapolated potential $V(r)$ at the physical point can then be used in the same way as potentials at non-physical pion masses, i.e.\ for fits of Eq.~(\ref{eq:ansatz}), using various $t$- and $r$-intervals to account for systematic uncertainties.

\begin{table}[t!]
\setlength{\tabcolsep}{0.16cm}
\begin{center}
  \caption{\label{tab:fitting}Extracted values of the fitting parameters $\alpha$ and $d$ (in fm) and of the binding energy $E$ (in MeV) in the scalar isosinglet channel ($I=0$, $j=0$). We show results for three ensembles differing in the pion mass and for the potentials extrapolated to the physical pion mass.}
\begin{tabular}{ccccc}
    Ensemble & $m_{\pi}$ [MeV] & $\alpha$ & $d$ [fm] & $E_B$ [MeV]\\
\hline
  B$150.24 $ & 650 & $0.31^{+0.03}_{-0.03}$ & $0.34^{+0.03}_{-0.03}$ & $-30^{+10}_{-12}$\\
  B$85.24 $  & 480 & $0.28^{+0.02}_{-0.02}$ & $0.37^{+0.04}_{-0.04}$ & $-27^{+9}_{-8}$\\
  B$40.24 $  & 340 & $0.35^{+0.04}_{-0.04}$ & $0.42^{+0.08}_{-0.08}$ & $-90^{+46}_{-42}$\\
  extrapolation & 140 & $0.34^{+0.03}_{-0.03}$ & $0.45^{+0.12}_{-0.10}$ & $-90^{+43}_{-36}$\\
  \end{tabular}
\end{center}
\end{table}

\begin{table}[t!]
\setlength{\tabcolsep}{0.16cm}
\begin{center}
  \caption{\label{tab:fitting_vector}Extracted values of the fitting parameters $\alpha$ and $d$ (in fm) in the vector isotriplet channel ($I=1$, $j=1$). No binding is observed. We show results for three ensembles differing in the pion mass and for the potentials extrapolated to the physical pion mass.}
\begin{tabular}{cccc}
    Ensemble & $m_{\pi}$ [MeV] & $\alpha$ & $d$ [fm]\\
\hline
  B$150.24 $ & 650 & $0.28^{+0.04}_{-0.04}$ & $0.15^{+0.02}_{-0.01}$\\
  B$85.24 $  & 480 & $0.30^{+0.06}_{-0.05}$ & $0.14^{+0.04}_{-0.02}$\\
  B$40.24 $  & 340 & $0.29^{+0.04}_{-0.06}$ & $0.16^{+0.03}_{-0.02}$\\
  extrapolation & 140 & $0.29^{+0.05}_{-0.06}$ & $0.16^{+0.05}_{-0.02}$\\
  \end{tabular}
\end{center}
\end{table}

\paragraph{Scalar isosinglet channel.} The results of this procedure for our different ensembles are shown in Tab.\ \ref{tab:fitting}, together with the outcome for the extrapolated potential.
The tendency that we observe is clear -- binding in the scalar isosinglet case ($I=0$, $j=0$) becomes stronger towards the physical pion mass. For the latter, we observe binding of:
\begin{equation}
E_B=-90^{+43}_{-36} {\rm \; MeV} . 
\end{equation}
Note that the result at the physical point is compatible with the one for ensemble B40.24, i.e.\ the pion mass dependence of the potential is relatively mild close to the physical point.
The conclusion that the attraction between two $B$ mesons (with a static antiquark and a light up/down quark of physical mass) in the $I=0$ channel is strong enough to form a tetraquark state and the value of the binding energy are among the main physical results of our paper.
Although they were obtained with only a single lattice spacing, we have strong hints that cut-off effects are under control and should not affect the final conclusion.

\paragraph{Vector isotriplet channel.} An analogous procedure for the vector isotriplet channel ($I=1$, $j=1$) gives the results in Tab.\ \ref{tab:fitting_vector}. Regardless of the pion mass, we observe no binding and the results are essentially independent on $m_\pi$ within our precision.
With these results, we agree with Ref.\ \cite{Ikeda:2013vwa}, where a similar study was done for $D\,D$ systems.
Moreover, in our analysis, the parameter $\alpha$ is the same for $I=0$ and $I=1$ potentials within uncertainties.
Hence, it is the much smaller value of the potential range $d$ that is responsible for the absence of binding in the vector isotriplet channel, as compared to the scalar isosinglet case.

\begin{figure}[t!]
\begin{center}
\includegraphics[width=0.99\columnwidth,angle=0]{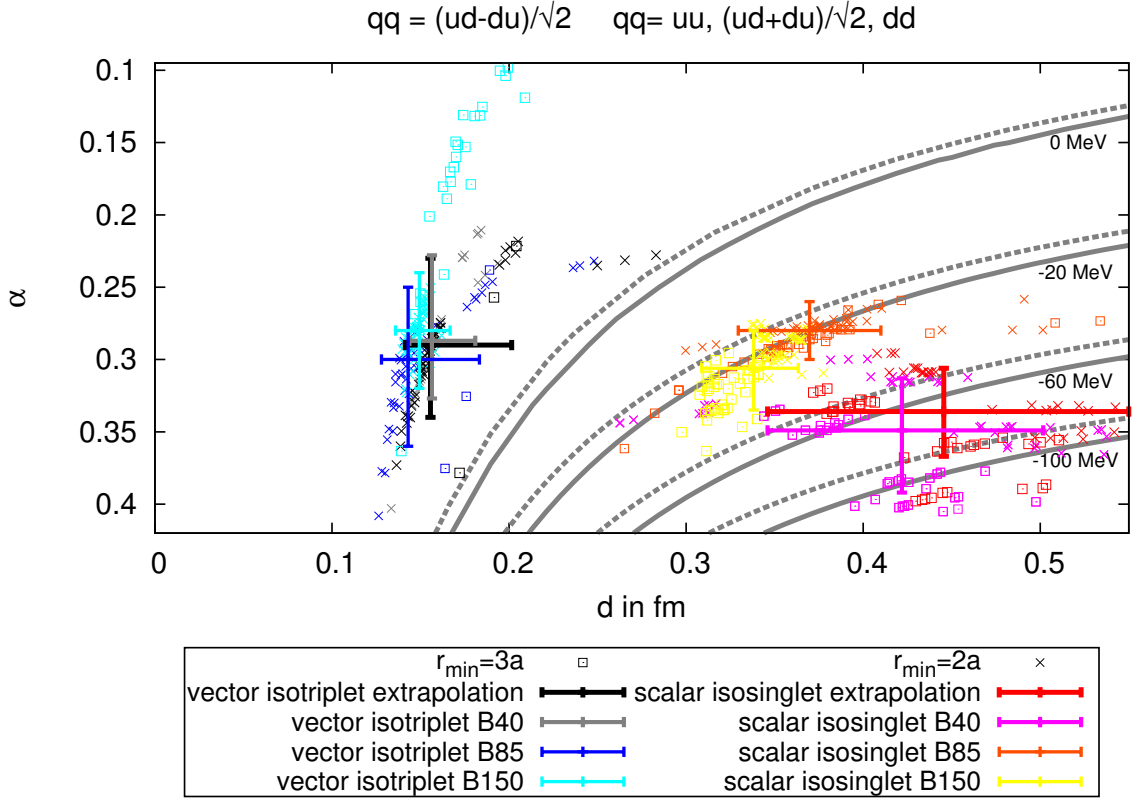}
\caption{\label{fig:summary} Binding energy isolines $E_B(\alpha,d)=\rm{const}$ in the $\alpha-d$ plane for the scalar isosinglet ($I=0$, $j=0$) and vector isotriplet ($I=1$, $j=1$) channels and four pion masses: 140 MeV (extrapolated), 340 MeV, 480 MeV and 650 MeV. The dashed and solid lines correspond to $m_H=m_B$ or $m_H=m_b$ in the Schr\"odinger equation, respectively. 
The crosses and squares are the fits of Eq.~(\ref{eq:ansatz}) for $r_{min}/a=2,3$, respectively, and different $r_{max}/a$ and $t$-intervals. The error bars represent combined systematic and statistical errors.
}
\end{center}
\end{figure}

\paragraph{Summary for both channels.} Finally, we summarize our results for both channels in Fig.\ \ref{fig:summary}. For each pion mass, we show the results of individual fits of Eq.~(\ref{eq:ansatz}) for different $t$- and $r$-fitting intervals, as well as the final error bar reflecting the combined statistical and systematic uncertainties. Values above or left of the binding threshold (the isoline \mbox{0 MeV}) correspond to no binding, while ones below or right of this threshold indicate that a bound state exists.
The central values of the error bars correspond to the respective entries in Tabs.~\ref{tab:fitting} and \ref{tab:fitting_vector}.


\section{Conclusions}
\label{sec:conclusions}

We have performed a lattice computation of potentials between two $B$ mesons in the static approximation.
We have investigated various channels, characterized by different quantum numbers $(I , I_z , |j_z| , \mathcal{P} , \mathcal{P}_x)$.
The most interesting cases from the point of view of formation of tetraquark states are the
attractive potentials with twice the mass of the $S$ meson as asymptotic value.
After identifying such potentials, they can be fitted by a phenomenologically motivated fitting ansatz and then inserted
in the Schr\"odinger equation to assess whether a bound state can form.
We have already followed such approach in our previous publication, however only at a single non-physical value of the pion mass.

In this work, our computations have been performed at three values of the pion mass -- from around 340 MeV to 650 MeV.
This has allowed for an extrapolation of the potentials to the physical pion mass, leading to one of the main conclusions of this paper: two $B$ mesons built of a static antiquark and a light quark of physical up/down quark mass form a tetraquark state in the scalar isosinglet channel.
The strength of the binding has been found to be $E_B=-90^{+43}_{-36} {\rm \; MeV}$, i.e.\ binding at a confidence level of more than $2\sigma$.

In addition to this result, we have shown in detail the results for the potentials at our lightest pion mass of approximately 340 MeV, i.e.\ a rather light dynamical quark mass, including for the first time also potentials between excited mesons.
Although our computations have been performed at a single value of the lattice spacing, we have found certain indication that the cut-off effects are not very large by analyzing different channels leading to the same result in the continuum limit.

One of our main findings is also a rule stating that a considered potential is attractive when the trial state
corresponding to the potential is symmetric under combined exchange of flavour, spin and parity.
We have also discussed a justification of this rule based on phenomenological considerations.

As a follow-up work, we plan to further investigate sources of systematic errors in our approach.
In particular, this includes a better control of cut-off effects, finite volume effects and quark mass effects. For the
latter, we plan to use ensembles at the physical pion mass, currently generated by the European Twisted Mass Collaboration.
It would also be an interesting direction to supplement the lattice computation by a perturbative
calculation of $B B$ potentials at small meson separations. 

We also plan to include effects due to the heavy quark spins following the exploratory study of Ref.\ \cite{Scheunert:2015pqa}. This is most important, since the contribution of the heavy quark spin to the hyperfine splitting is of the some order as the binding energy we obtain for the tetraquark.


\section*{Acknowledgments}
P.B. thanks IFT for hospitality and CFTP, grant FCT UID/FIS/00777/2013, for support. 
M.W. and A.P. acknowledge support by the Emmy Noether Programme of the DFG (German Research Foundation), grant WA 3000/1-1.
This work was supported in part by the Helmholtz International Center for FAIR within the framework of the LOEWE program launched by the State of Hesse.
Calculations on the LOEWE-CSC high-performance computer of Johann Wolfgang Goethe-University Frankfurt am Main were conducted for this research. We would
like to thank HPC-Hessen, funded by the State Ministry of Higher Education, Research and the Arts, for programming advice.



\begin{thebibliography}{99}
\bibitem{Jaffe:1976ig}
  R.~L.~Jaffe,
  Phys.\ Rev.\ D {\bf 15}, 267 (1977).

\bibitem{Jaffe:2004ph}
  R.~L.~Jaffe,
  Phys.\ Rept.\ {\bf 409}, 1 (2005) 
  [arXiv: hep-ph/0409065].
  
\bibitem{Lipkin:1987sk}
  H.~J.~Lipkin,
  Phys.\ Lett.\ B {\bf 195}, 484 (1987).


\bibitem{Praszalowicz:2003ik}
  M.~Praszalowicz,
  Phys.\ Lett.\ B {\bf 575}, 234 (2003)
  [arXiv: hep-ph/0308114].
  
\bibitem{Aaij:2015tga}
  R.~Aaij {\it et al.} [LHCb Collaboration],
  Phys.\ Rev.\ Lett.\  {\bf 115}, 072001 (2015)
  [arXiv:1507.03414 [hep-ex]].
  
\bibitem{Liu:2013dau}
  Z.~Q.~Liu {\it et al.} [Belle Collaboration],
  Phys.\ Rev.\ Lett.\ {\bf 110}, 252002 (2013)
  [arXiv:1304.0121 [hep-ex]].

\bibitem{Chilikin:2014bkk}
  K.~Chilikin {\it et al.} [Belle Collaboration],
  Phys.\ Rev.\ D {\bf 90}, 112009 (2014)
  [arXiv:1408.6457 [hep-ex]].

\bibitem{Xiao:2013iha}
  T.~Xiao, S.~Dobbs, A.~Tomaradze and K.~K.~Seth,
  Phys.\ Lett.\ B {\bf 727}, 366 (2013)
  [arXiv:1304.3036 [hep-ex]].

\bibitem{Ablikim:2013mio}
  M.~Ablikim {\it et al.} [BESIII Collaboration],
  Phys.\ Rev.\ Lett.\ {\bf 110}, 252001 (2013)
  [arXiv:1303.5949 [hep-ex]].

\bibitem{Ablikim:2013emm}
  M.~Ablikim {\it et al.} [BESIII Collaboration],
  Phys.\ Rev.\ Lett.\ {\bf 112}, 132001 (2014)
  [arXiv:1308.2760 [hep-ex]].

\bibitem{Ablikim:2013wzq} 
  M.~Ablikim {\it et al.} [BESIII Collaboration],
  Phys.\ Rev.\ Lett.\ {\bf 111}, 242001 (2013)
  [arXiv:1309.1896 [hep-ex]].

\bibitem{Ablikim:2013xfr}
  M.~Ablikim {\it et al.} [BESIII Collaboration],
  Phys.\ Rev.\ Lett.\ {\bf 112}, 022001 (2014)
  [arXiv:1310.1163 [hep-ex]].

\bibitem{Ablikim:2014dxl}
  M.~Ablikim {\it et al.} [BESIII Collaboration],
  Phys.\ Rev.\ Lett.\ {\bf 113}, 212002 (2014)
  [arXiv:1409.6577 [hep-ex]].

\bibitem{Aaij:2014jqa}
  R.~Aaij {\it et al.} [LHCb Collaboration],
  Phys.\ Rev.\ Lett.\ {\bf 112}, 222002 (2014)
  [arXiv:1404.1903 [hep-ex]].

\bibitem{Prelovsek:2010kg}
  S.~Prelovsek, T.~Draper, C.~B.~Lang, M.~Limmer, K.~F.~Liu, N.~Mathur and D.~Mohler,
  Phys.\ Rev.\ D {\bf 82}, 094507 (2010)
  [arXiv:1005.0948 [hep-lat]].
  
\bibitem{Alexandrou:2012rm}
  C.~Alexandrou, J.~O.~Daldrop, M.~Dalla Brida, M.~Gravina, L.~Scorzato, C.~Urbach and M.~Wagner,
  JHEP {\bf 1304}, 137 (2013)
  [arXiv:1212.1418 [hep-lat]].

  \bibitem{Mohler:2013rwa}
  D.~Mohler, C.~B.~Lang, L.~Leskovec, S.~Prelovsek and R.~M.~Woloshyn,
  Phys.\ Rev.\ Lett.\  {\bf 111},  222001 (2013)
  [arXiv:1308.3175 [hep-lat]].  
  
  
\bibitem{Prelovsek:2014swa}
  S.~Prelovsek, C.~B.~Lang, L.~Leskovec and D.~Mohler,
  Phys.\ Rev.\ D {\bf 91},  014504 (2015)
  [arXiv:1405.7623 [hep-lat]].  
  
\bibitem{Abdel-Rehim:2014zwa}
  J.~Berlin, A.~Abdel-Rehim, C.~Alexandrou, M.~Dalla Brida, M.~Gravina and M.~Wagner,
  PoS LATTICE {\bf 2014} 104 (2014) 
  [arXiv:1410.8757 [hep-lat]].

\bibitem{Berlin:2015faa}
  J.~Berlin, A.~Abdel-Rehim, C.~Alexandrou, M.~D.~Brida, M.~Gravina and M.~Wagner,
  arXiv:1508.04685 [hep-lat].  
  
  
  
  
\bibitem{Guerrieri:2014nxa}
  A.~L.~Guerrieri {\it et al.},
  PoS LATTICE {\bf 2014}, 106 (2014)
  [arXiv:1411.2247 [hep-lat]].

  
\bibitem{Ikeda:2013vwa} 
  Y.~Ikeda {\it et al.},
  Phys.\ Lett.\ B {\bf 729}, 85 (2014)
  [arXiv:1311.6214 [hep-lat]].
  

\bibitem{Alexandrou:2004ak}
  C.~Alexandrou and G.~Koutsou,
  Phys.\ Rev.\ D {\bf 71}, 014504 (2005)
  [hep-lat/0407005].

\bibitem{Okiharu:2004ve} 
  F.~Okiharu, H.~Suganuma and T.~T.~Takahashi,
  Phys.\ Rev.\ D {\bf 72}, 014505 (2005)
  [hep-lat/0412012].
  
\bibitem{Cardoso:2011fq}
  N.~Cardoso, M.~Cardoso and P.~Bicudo,
  Phys.\ Rev.\ D {\bf 84}, 054508 (2011)
  [arXiv:1107.1355 [hep-lat]].

\bibitem{Cardoso:2012uka}
  M.~Cardoso, N.~Cardoso and P.~Bicudo,
  Phys.\ Rev.\ D {\bf 86}, 014503 (2012)
  [arXiv:1204.5131 [hep-lat]].
  
\bibitem{Bicudo:2015bra} 
  P.~Bicudo and M.~Cardoso,
  arXiv:1509.04943 [hep-ph].
  
  
  
\bibitem{Bicudo:2012qt}
  P.~Bicudo and M.~Wagner,
  Phys.\ Rev.\ D {\bf 87}, 114511 (2013)
  [arXiv:1209.6274 [hep-ph]].

\bibitem{Brown:2012tm}
  Z.~S.~Brown and K.~Orginos,
  Phys.\ Rev.\ D {\bf 86}, 114506 (2012)
  [arXiv:1210.1953 [hep-lat]].
  
\bibitem{Bicudo:2015vta}
  P.~Bicudo, K.~Cichy, A.~Peters, B.~Wagenbach and M.~Wagner,
  Phys.\ Rev.\ D {\bf 92}, 014507 (2015)
  [arXiv:1505.00613 [hep-lat]].

\bibitem{Belle:2011aa}
  A.~Bondar {\it et al.} [Belle Collaboration],
  Phys.\ Rev.\ Lett.\ {\bf 108}, 122001 (2012)
  [arXiv:1110.2251 [hep-ex]].  
  
\bibitem{Ader:1981db} 
  J.~P.~Ader, J.~M.~Richard and P.~Taxil,
  Phys.\ Rev.\ D {\bf 25}, 2370 (1982).
  
\bibitem{Ballot:1983iv}
  J.~L.~Ballot and J.~M.~Richard,
  Phys.\ Lett.\ B {\bf 123}, 449 (1983).
  
\bibitem{Heller:1986bt} 
  L.~Heller and J.~A.~Tjon,
  Phys.\ Rev.\ D {\bf 35}, 969 (1987).

\bibitem{Carlson:1987hh} 
  J.~Carlson, L.~Heller and J.~A.~Tjon,
  Phys.\ Rev.\ D {\bf 37}, 744 (1988).

\bibitem{Lipkin:1986dw} 
  H.~J.~Lipkin,
  Phys.\ Lett.\ B {\bf 172}, 242 (1986).

\bibitem{Brink:1998as} 
  D.~M.~Brink and F.~Stancu,
  Phys.\ Rev.\ D {\bf 57}, 6778 (1998).

\bibitem{Gelman:2002wf} 
  B.~A.~Gelman and S.~Nussinov,
  Phys.\ Lett.\ B {\bf 551}, 296 (2003)
  [arXiv: hep-ph/0209095].

\bibitem{Vijande:2003ki} 
  J.~Vijande, F.~Fernandez, A.~Valcarce and B.~Silvestre-Brac,
  Eur.\ Phys.\ J.\ A {\bf 19}, 383 (2004)
  [arXiv: hep-ph/0310007].

\bibitem{Janc:2004qn} 
  D.~Janc and M.~Rosina,
  Few Body Syst.\  {\bf 35}, 175 (2004)
  [arXiv: hep-ph/0405208].

\bibitem{Cohen:2006jg} 
  T.~D.~Cohen and P.~M.~Hohler,
  Phys.\ Rev.\ D {\bf 74}, 094003 (2006)
  [arXiv: hep-ph/0606084].

\bibitem{Vijande:2007ix} 
  J.~Vijande, A.~Valcarce and J.-M.~Richard,
  Phys.\ Rev.\ D {\bf 76}, 114013 (2007)
  [arXiv:0707.3996 [hep-ph]].
 

\bibitem{Born:1927}
  M.~Born and R.~Oppenheimer,
  Annalen der Physik \textbf{389}, 457 (1927).
  
  
  
 
  
\bibitem{Richards:1990xf} 
  D.~G.~Richards, D.~K.~Sinclair and D.~W.~Sivers,
  Phys.\ Rev.\ D {\bf 42}, 3191 (1990).
  
\bibitem{Mihaly:1996ue} 
  A.~Mihaly, H.~R.~Fiebig, H.~Markum and K.~Rabitsch,
  Phys.\ Rev.\ D {\bf 55}, 3077 (1997).
  
\bibitem{Stewart:1998hk} 
  C.~Stewart and R.~Koniuk,
  Phys.\ Rev.\ D {\bf 57}, 5581 (1998)
  [arXiv: hep-lat/9803003].
    
\bibitem{Michael:1999nq} 
  C.~Michael and P.~Pennanen [UKQCD Collaboration],
  Phys.\ Rev.\ D {\bf 60}, 054012 (1999)
  [arXiv: hep-lat/9901007].
  
      
\bibitem{Cook:2002am}
  M.~S.~Cook and H.~R.~Fiebig,
  arXiv: hep-lat/0210054.

\bibitem{Doi:2006kx}
  T.~Doi, T.~T.~Takahashi and H.~Suganuma,
  AIP Conf.\ Proc.\ {\bf 842}, 246 (2006)
  [arXiv: hep-lat/0601008].

\bibitem{Detmold:2007wk}
  W.~Detmold, K.~Orginos and M.~J.~Savage,
  Phys.\ Rev.\ D {\bf 76}, 114503 (2007)
  [arXiv: hep-lat/0703009].

\bibitem{Wagner:2010ad} 
  M.~Wagner [ETM Collaboration],
  PoS LATTICE {\bf 2010}, 162 (2010)
  [arXiv:1008.1538 [hep-lat]].

\bibitem{Bali:2010xa}
  G.~Bali and M.~Hetzenegger,
  PoS {\bf LATTICE2010}, 142 (2010)
  [arXiv: 1011.0571 [hep-lat]].
  
\bibitem{Wagner:2011ev} 
  M.~Wagner [ETM Collaboration],
  Acta Phys.\ Polon.\ Supp.\  {\bf 4}, 747 (2011)
  [arXiv:1103.5147 [hep-lat]].

\bibitem{Ribeiro:1978gx} 
  J.~E.~T.~Ribeiro,
  Z.\ Phys.\ C {\bf 5}, 27 (1980).

\bibitem{Myhrer:1987af} 
  F.~Myhrer and J.~Wroldsen,
  Rev.\ Mod.\ Phys.\  {\bf 60}, 629 (1988).
  
\bibitem{Downum:2010qa} 
  C.~Downum, J.~Stone, T.~Barnes, E.~Swanson and I.~Vidana,
  AIP Conf.\ Proc.\  {\bf 1257}, 538 (2010)
  [arXiv: 1001.3320 [nucl-th]].
  
\bibitem{Ordonez:1993tn} 
  C.~Ordonez, L.~Ray and U.~van Kolck,
  Phys.\ Rev.\ Lett.\  {\bf 72}, 1982 (1994).
  
\bibitem{Kaiser:1997mw} 
  N.~Kaiser, R.~Brockmann and W.~Weise,
  Nucl.\ Phys.\ A {\bf 625}, 758 (1997)
  [arXiv: nucl-th/9706045].
  
\bibitem{Kaplan:1998tg} 
  D.~B.~Kaplan, M.~J.~Savage and M.~B.~Wise,
  Phys.\ Lett.\ B {\bf 424}, 390 (1998)
  [arXiv: nucl-th/9801034].
    
\bibitem{Epelbaum:2004fk} 
  E.~Epelbaum, W.~Glockle and U.~G.~Meissner,
  Nucl.\ Phys.\ A {\bf 747}, 362 (2005)
  [arXiv: nucl-th/0405048].
  

\bibitem{Frezzotti:2000nk}
  R.~Frezzotti, P.~A.~Grassi, S.~Sint and P.~Weisz,
  JHEP {\bf 0108}, 058 (2001)
  [arXiv: hep-lat/0101001].

\bibitem{Frezzotti:2003ni}
  R.~Frezzotti and G.~C.~Rossi,
  JHEP {\bf 0408}, 007 (2004)
  [arXiv: hep-lat/0306014].

\bibitem{Weisz:1982zw}
  P.~Weisz,
  Nucl.\ Phys.\ B {\bf 212}, 1 (1983).

\bibitem{Farchioni:2004ma}
F.~Farchioni {\it et al.}, 
  Nucl.Phys.Proc.Suppl. {\bf 140}, 240 (2005)
  [arXiv: hep-lat/0409098].

\bibitem{Farchioni:2004fs}
F.~Farchioni {\it et al.}, 
  Eur.Phys.J. C {\bf 42}, 73 (2005)
  [arXiv: hep-lat/0410031].

\bibitem{Frezzotti:2005gi}
R.~Frezzotti, G.~Martinelli, M.~Papinutto, and G.~Rossi, 
  JHEP {\bf 0604}, 038 (2006)
  [arXiv: hep-lat/0503034].

\bibitem{Jansen:2005kk}
K.~Jansen, M.~Papinutto, A.~Shindler, C.~Urbach, and
  I.~Wetzorke [XLF Collaboration],
  JHEP {\bf 0509}, 071 (2005)
  [arXiv: hep-lat/0507010].

\bibitem{Baron:2009wt}
  R.~Baron {\it et al.} [ETM Collaboration],
  JHEP {\bf 1008}, 097 (2010)
  [arXiv: 0911.5061 [hep-lat]].
  
\bibitem{Boucaud:2007uk}
  Ph.~Boucaud {\it et al.} [ETM Collaboration],
  Phys.\ Lett.\  B {\bf 650}, 304 (2007)
  [arXiv: hep-lat/0701012].

\bibitem{Boucaud:2008xu}
  P.~Boucaud {\it et al.} [ETM Collaboration],
  Comput.\ Phys.\ Commun.\ {\bf 179}, 695 (2008)
  [arXiv: 0803.0224 [hep-lat]].

\bibitem{Blossier:2010cr}
  B.~Blossier {\it et al.} [ETM Collaboration],
  Phys.\ Rev.\ D {\bf 82}, 114513 (2010)
  [arXiv: 1010.3659 [hep-lat]].    

\bibitem{Jansen:2011vv}
  K.~Jansen {\it et al.} [ETM Collaboration],
  JHEP {\bf 1201}, 025 (2012)
  [arXiv: 1110.6859 [hep-ph]].  
  


\bibitem{PDG}
 K.A. Olive et al. (Particle Data Group), Chinese Physics {\bf C38}, 090001 (2014).


\bibitem{Eichten:1987xu}
  E.~Eichten,
  Nucl.\ Phys.\ Proc.\ Suppl.\  {\bf 4}, 170 (1988). 
 

\bibitem{Eichten:1989zv}
  E.~Eichten and B.~R.~Hill,
  Phys.\ Lett.\ B {\bf 234}, 511 (1990).
  
\bibitem{Scheunert:2015pqa}
  J.~Scheunert, P.~Bicudo, A.~Uenver and M.~Wagner,
  Acta Phys.\ Polon.\ Supp.\  {\bf 8},  363 (2015)
  [arXiv: 1505.03496 [hep-ph]].  
  

\bibitem{Kalinowski:2015bwa}
  M.~Kalinowski and M.~Wagner,
  arXiv: 1509.02396 [hep-lat].
  
  
\bibitem{Abazov:2007vq}
  V.~M.~Abazov {\it et al.}  [D0 Collaboration],
  Phys.\ Rev.\ Lett.\  {\bf 99}, 172001 (2007)
  [arXiv: 0705.3229 [hep-ex]].

\bibitem{Aaltonen:2007ah}
  T.~Aaltonen {\it et al.}  [CDF Collaboration],
  Phys.\ Rev.\ Lett.\  {\bf 100}, 082001 (2008)
  [arXiv: 0710.4199 [hep-ex]].


\bibitem{Jansen:2008si}
  K.~Jansen, C.~Michael, A.~Shindler and M.~Wagner [ETM Collaboration],
  JHEP {\bf 0812}, 058 (2008)
  [arXiv: 0810.1843 [hep-lat]].

\bibitem{Michael:2010aa}
  C.~Michael, A.~Shindler and M.~Wagner [ETM Collaboration],
  JHEP {\bf 1008}, 009 (2010)
  [arXiv: 1004.4235 [hep-lat]].





\bibitem{Albanese:1987ds}
  M.~Albanese {\it et al.}  [APE Collaboration],
  Phys.\ Lett.\ B {\bf 192}, 163 (1987).


\bibitem{Gusken:1989qx}
  S.~Gusken,
  Nucl.\ Phys.\ Proc.\ Suppl.\  {\bf 17}, 361 (1990).

\bibitem{Wagner:2011fs}
  M.~Wagner and C.~Wiese [ETM Collaboration],
  JHEP {\bf 1107}, 016 (2011)
  [arXiv: 1104.4921 [hep-lat]].
  
  

  
  
  
  
\bibitem{Hasenfratz:2001hp}
  A.~Hasenfratz and F.~Knechtli,
  Phys.\ Rev.\ D {\bf 64}, 034504 (2001)
  [arXiv: hep-lat/0103029].


\bibitem{DellaMorte:2005yc}
  M.~Della Morte, A.~Shindler and R.~Sommer,
  JHEP {\bf 0508}, 051 (2005)
  [arXiv: hep-lat/0506008].


\bibitem{Blossier:2009kd}
  B.~Blossier, M.~Della Morte, G.~von Hippel, T.~Mendes and R.~Sommer,
  JHEP {\bf 0904}, 094 (2009)
  [arXiv: 0902.1265 [hep-lat]].

  
  
\bibitem{Xiao:2013yca} 
  C.~W.~Xiao, J.~Nieves and E.~Oset,
  Phys.\ Rev.\ D {\bf 88}, 056012 (2013)
  [arXiv: 1304.5368 [hep-ph]].


\bibitem{Ozpineci:2013qza} 
  A.~Ozpineci, C.~W.~Xiao and E.~Oset,
  Phys.\ Rev.\ D {\bf 88}, 034018 (2013)
  [arXiv: 1306.3154 [hep-ph]].
  
\bibitem{Torres:2013lka} 
  A.~Martinez Torres, K.~P.~Khemchandani, F.~S.~Navarra, M.~Nielsen and E.~Oset,
  Phys.\ Rev.\ D {\bf 89}, 014025 (2014)
  [arXiv: 1310.1119 [hep-ph]].

\bibitem{Aceti:2014kja} 
  F.~Aceti, M.~Bayar, J.~M.~Dias and E.~Oset,
  Eur.\ Phys.\ J.\ A {\bf 50}, 103 (2014)
  [arXiv: 1401.2076 [hep-ph]].

\bibitem{Aceti:2014uea} 
  F.~Aceti {\it et al.},
  Phys.\ Rev.\ D {\bf 90}, 016003 (2014)
  [arXiv: 1401.8216 [hep-ph]].



\bibitem{Dias:2014pva} 
  J.~M.~Dias, F.~Aceti and E.~Oset,
  Phys.\ Rev.\ D {\bf 91}, 076001 (2015)
  [arXiv: 1410.1785 [hep-ph]].
  
\bibitem{Barnes:1999hs} 
  T.~Barnes, N.~Black, D.~J.~Dean and E.~S.~Swanson,
  Phys.\ Rev.\ C {\bf 60}, 045202 (1999)
  [arXiv: nucl-th/9902068].
  

  
  
  
\end{thebibliography}
\end{document}